\documentclass[aps,pre,reprint,superscriptaddress,amsmath,amssymb,longbibliography,floatfix]{revtex4-2}

\usepackage{graphicx}
\usepackage{bm}
\usepackage{booktabs}
\usepackage{array}
\usepackage{xurl}
\usepackage[hidelinks]{hyperref}

\begin{document}

\title{Replacement distance controls growth at long wavelengths in a coevolving spatial network}

\author{Ruiwu Niu}
\email[Corresponding author: ]{rniu@hksyu.edu}
\affiliation{Department of Data Science and Digital Innovation, Hong Kong Shue Yan University, Hong Kong SAR, China}
\author{Ming Yuan}
\email[Email: ]{yuanming@sjtu.edu.cn}
\affiliation{State Key Laboratory of Mechanical System and Vibration, Shanghai Jiao Tong University, Shanghai 200240, China}
\author{Ying-kit Wong}
\email[Email: ]{ykwong@hksyu.edu}
\affiliation{Department of Data Science and Digital Innovation, Hong Kong Shue Yan University, Hong Kong SAR, China}

\date{October 12, 2026}

\begin{abstract}
When an agent leaves a spatial network, its replacement removes old contacts
and may transport population to a new location. The frequency of replacement
alone therefore does not determine how turnover changes spatial organization.
We study diffusing agents joined by persistent attractive contacts, with the
total pull on each agent bounded. At fixed population, successors appear at
the departure position, within a surrounding disk, or anywhere in the system.
We ask whether small differences between dense and dilute regions grow or
fade. The theory separates an exactly calculable diffusion and replacement
contribution from an approximate attraction contribution obtained by averaging
the contact network. Paired simulations find similar attraction contributions
under the three rules at all nine sampled wavelengths. Changing only the
replacement location reverses the broadest tested pattern: it grows with
replacement at the departure position but fades under both displaced rules.
An independent radius scan traces this change from growth to decay.
Under displaced replacement, intermediate wavelengths grow between sampled
long and short scales with decay. Rates measured in
the earliest finite window are compatible with the prediction at long and
intermediate wavelengths; discrepancies are already present at short
wavelengths. Replacement distance thus helps determine which spatial density
contrasts are initially amplified.
\end{abstract}

\maketitle

\section{Introduction}
\label{sec:introduction}

Does it matter where a new agent appears when another leaves a spatial
network? The replacement rate says how often this occurs, but not where
population moves. Replacement at the departure position preserves the
instantaneous density; replacement elsewhere transports it. Systems with
the same population and replacement frequency can therefore have different
spatial dynamics.

Persistent contacts make this distinction important. Agents can retain
interaction partners after separating, just as particles can remain joined
by temporary bonds. Temporal networks describe the changing set of contacts
\cite{HolmeSaramaki2012Temporal}; when contacts also affect node dynamics,
the resulting feedback makes the network adaptive
\cite{GrossBlasius2008Adaptive,Berner2023AdaptiveDynamical}.
Here encounters create edges that alter subsequent motion.
Attractive edges can draw agents together and promote further encounters.
A departure interrupts this feedback by removing incident edges. Relocation
also transports density. We ask whether changing this second effect, while
keeping replacement frequency and edge removal fixed, changes which spatial
patterns grow.

Here a pattern is a variation between denser and more dilute regions.
A broad pattern spans much of the system; a finer one contains several
smaller regions. Growth means increasing density contrast, not increasing
population. Short displacements barely affect a broad variation, whereas
longer displacements mix its dense and dilute regions. Diffusion smooths
finer variations most strongly, while attraction can amplify contrast.
Comparing these processes across wavelengths connects replacement to
the initial stages
of pattern formation \cite{CrossHohenberg1993PatternFormation}.

Spatial replacement is closely related to stochastic resetting. Returning a
diffusing particle to a prescribed location changes its stationary
distribution and first passage statistics
\cite{EvansMajumdar2011Resetting,EvansMajumdarSchehr2020Resetting}, and resetting
has also been studied for interacting particles
\cite{NagarGupta2023InteractingResetting,VatashRoichman2025ManyBodyResetting}.
Random reset amplitudes extend resetting to distributed displacements
\cite{Dahlenburg2021RandomAmplitude}, and density equations have been derived
for interacting particles with independent or simultaneous resets
\cite{Bressloff2024DensityResetting}.
We draw replacement displacements from the departing agent's current
position. These translation invariant rules are specified by a displacement
distribution and also remove the departing agent's contacts.

Mobile encounter models connect accumulated contacts to network structure
\cite{Manna2003Friendship,Gonzalez2006MobileAgents}, while attractiveness and
similarity help explain contact duration and recurrence
\cite{Starnini2013FaceToFace,RodriguezFlores2018Similarity}.
Node activity and spatial proximity offer further mechanisms for temporal
contacts and their correlations
\cite{Perra2012ActivityDriven,HartleMasuda2025Autocorrelation,SimonSaramaki2026Spatiotemporal}.
Hysteretic springs model bonds that form on contact and persist until a
larger separation is reached \cite{Krinninger2016DynamicBonding}.
In linked colloidal networks, bond persistence also governs structural
relaxation \cite{Kwon2022LinkedNetworks}.
Here this coupling between persistent contacts and motion provides the
attraction against which replacement transport competes.

Other transport and population processes can also control spatial scales.
Nonlocal competition can produce periodic clusters
\cite{Fuentes2003Nonlocal,HernandezGarcia2004Clustering}, and chemical reactions
can suppress broad composition variations in phase separating mixtures
\cite{Glotzer1995Reaction}. Birth and death can arrest motility driven
separation or sustain cycles of colony formation and dispersal
\cite{Cates2010Arrested,Grafke2017Spatiotemporal}; they can also alter
structural order in Brownian particle systems \cite{Almodovar2022BirthDeath}.
Active currents and reactions provide routes to microphase separation,
in which domains remain finite
\cite{Tjhung2018Cluster,LiCates2021Hierarchical}.
In reaction diffusion and chemotactic systems, production and degradation
can interrupt the growth of domains and set the scales of developed patterns
\cite{Brauns2021Wavelength,Weyer2023Coarsening,Weyer2025Chemotaxis}.
Brownian birth and death models can cluster without attraction when offspring are born
beside existing organisms, creating reproductive pair correlations
\cite{Young2001Reproductive,Houchmandzadeh2002Clustering}.
Our population is fixed, with exactly one successor per departure.
Replacement at the same position preserves the spatial configuration;
persistent edges supply the attraction in our model.

We isolate this competition in a model of diffusing agents with reciprocal
attractive contacts. Three replacement rules put each successor at the
departure position, uniformly within a surrounding disk, or uniformly over
the whole system. All three erase the departing agent's edges. We normalize
each interaction by the strengths of both endpoints so that an agent's
total attractive drift remains bounded for every graph and system size.
Increasing the number of contacts therefore cannot make the pull grow
without bound.

Without attraction, diffusion and replacement give an exactly calculable
rate for each density wave. Averaging the contact network, using the contact
statistics of a pair of independently diffusing agents as reference, gives
an approximate attraction contribution. Paired simulations test this contribution, the exact baseline,
and their sum separately. This separates transport accuracy from the
accuracy of the network approximation. It also exposes cancellation:
smoothing and attraction can be individually strong but leave a small
total rate, so modest errors in attraction can substantially change that
remainder.

Across nine sampled wavelengths, the attraction contribution is similar
under the three replacement rules. Changing where replacements appear
reverses the broadest tested density pattern: it grows under replacement
at the departure position and fades under either displaced rule. An
independent radius scan traces this change from growth to decay. Additional
modes resolve intermediate growth between sampled scales with decay under
the displaced rules.

\begin{figure*}[t!]
  \centering
  \includegraphics[width=\textwidth]{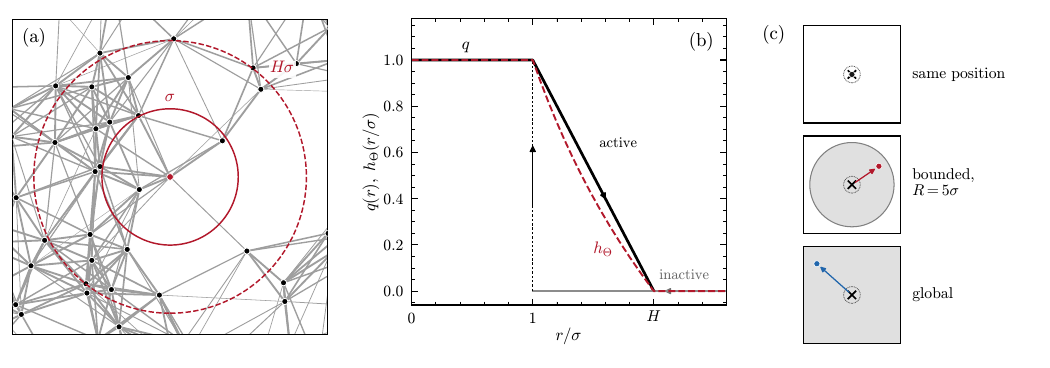}
  \caption{\textbf{Microscopic rules of the coevolving spatial network.}
  (a) A window of side $0.4L$ from one initial configuration generated by the
  simulator with $N=256$, $\eta=6$, $H=2$, and $\Theta=5$. This sample
  illustrates the initialization, not an evolved state. Gray lines are
  active edges; their widths increase with the edge weight $q$.
  The solid and dashed circles around the highlighted agent have radii
  $\sigma$ and $H\sigma$, respectively.
  (b) Hysteretic edge rule. An inactive pair follows the gray branch with
  zero coupling and becomes active at $r\leq\sigma$ (vertical arrow).
  The black branch gives the weight $q(r)$ of an active edge, which is
  deleted at $r=H\sigma$. The red dashed curve is the conditional edge
  activity probability $h_\Theta(r/\sigma)$ in the reference state without
  attraction [Eq.~\eqref{eq:pair-profile}]; it is a probability, not an edge weight.
  (c) Three replacement rules, drawn to scale within separate copies of the
  full periodic square. A departing agent at the cross is replaced at the
  same position, uniformly within a disk of radius $R=5\sigma$ (shaded), or
  uniformly over the whole domain (shaded). Dots indicate example arrival
  positions, and dotted circles have radius $\sigma$. All three rules erase
  the departing agent's incident edges.}
  \label{fig:model}
\end{figure*}

The accuracy of the averaged theory depends on scale and observation time.
Rates over the earliest finite window are compatible with its prediction
at the tested long and intermediate wavelengths; both short modes already
show a discrepancy. We measure the early response to a prepared weak
density modulation. After introducing the microscopic rules in
Fig.~\ref{fig:model}, we derive their transport contribution and test the
contact approximation and total response separately.

\section{Coevolving spatial network with persistent contact edges}
\label{sec:model}

Figure~\ref{fig:model} connects the model's three ingredients. Panel (a)
shows an initial spatial network on a periodic square, where leaving one
side means entering the opposite side. Panel (b) shows how different contact
and deletion distances preserve encounter history. Its dashed curve previews
the contact probability derived in Sec.~\ref{sec:persistence}. Panel (c)
compares replacement locations at their actual scale. The rules below make
explicit how motion changes the graph and how its edges feed back on motion.

\subsection{Motion, persistent contacts, and force}

Let \(N\in\{2,3,\ldots\}\) agents occupy a periodic square
\(\mathbb T_L^2=[0,L)^2\), where \(L>0\) and opposite sides are identified.
At physical time \(t_{\rm phys}\in[0,\infty)\), agent
\(i\in\{1,\ldots,N\}\) has position
\(X_i(t_{\rm phys})\in\mathbb T_L^2\). Persistent computational labels keep the
population exactly equal to \(N\). At rate \(\mu>0\), the agent at a label is
removed and immediately replaced by a new agent. The label serves only as
bookkeeping; the departing agent and all its incident edges are removed.
Periodic boundaries remove wall effects and let us focus on aggregation
created by the agents. For distinct indices \(i\) and \(j\), the displacement
\(\Delta_{ij}\in[-L/2,L/2)^2\) is the shortest periodic vector from agent
\(i\) to agent \(j\), and \(r_{ij}=|\Delta_{ij}|\in[0,L/\sqrt2]\) is their
separation.

A contact stores one bit of history: whether the pair has recently met. An
unconnected pair forms an edge when it first comes within the contact distance
\(\sigma>0\). The edge is not deleted as soon as the pair crosses that same
distance. Instead, it survives until the separation reaches the larger value
\(H\sigma\), with \(H>1\). We assume \(H\sigma<L/2\), as in all simulations.
This use of different formation and deletion
thresholds is called \emph{hysteresis}. While the edge is active, its dimensionless
weight is

\begin{equation}
\label{eq:weight}
q(r)=
\begin{cases}
1, & r\leq\sigma,\\[2pt]
\dfrac{H-r/\sigma}{H-1}, & \sigma<r<H\sigma,\\[6pt]
0, & r\geq H\sigma.
\end{cases}
\end{equation}

The weight is one inside the contact distance, decreases linearly during
separation, and recovers if the active pair approaches again. A disconnected
pair at the same intermediate distance remains disconnected until it crosses
the inner threshold. Thus distance alone does not determine the graph: two
pairs at the same separation can have different edge states because they have
different encounter histories. The function
\(q:[0,\infty)\to[0,1]\) is dimensionless.

Let \(A_{ij}(t_{\rm phys})\in\{0,1\}\) indicate whether the undirected edge
\(ij\) is active, with \(A_{ij}=A_{ji}\) and \(A_{ii}=0\). The symmetric
matrix \(A(t_{\rm phys})=(A_{ij}(t_{\rm phys}))\) is the active contact graph;
we omit its explicit time argument below. Let
\(q_{ij}=q(r_{ij})\), and
\(s_i=\sum_{j\ne i} A_{ij}q_{ij}\geq0\) the weighted degree of agent \(i\).
Let \(\kappa>0\) be the diffusion coefficient, let \(\chi\geq0\) set the
attraction rate, let \(s_0>0\) be a regularizing strength, and let
\(B_i(t_{\rm phys})\in\mathbb R^2\) be independent standard Brownian motions.
Between replacement events, the agent positions obey

\begin{equation}
\label{eq:motion}
\begin{aligned}
dX_i={}&
\chi\sum_{j\ne i}
\frac{A_{ij}q_{ij}}{s_0+s_i+s_j}\,\Delta_{ij}\,dt_{\rm phys}\\
&+\sqrt{2\kappa}\,dB_i(t_{\rm phys})
\pmod L.
\end{aligned}
\end{equation}

The final term is ordinary Brownian motion. The first term is the sum of
the pulls from active contacts, weighted by their current strengths. The
constant \(s_0\) keeps the denominator finite for isolated or weakly
connected agents. Because an edge has the same denominator when its endpoints
are exchanged, it gives equal and opposite drift increments to the two agents.
Agents are points: the model includes neither excluded volume nor a repulsive
force at short separation.

\subsection{Why the attraction is normalized}

The normalization in Eq.~\eqref{eq:motion} separates the effect of replacement
distance from the increasing pull that an unnormalized force sum would
produce as contacts accumulate. Let
\(b_i\in\mathbb R^2\) denote the attractive drift, namely the coefficient of
\(dt_{\rm phys}\) in the first term of Eq.~\eqref{eq:motion}. Because every active edge satisfies
\(r_{ij}<H\sigma\), it obeys the exact bound

\begin{equation}
\label{eq:drift-bound}
|b_i|\leq\chi H\sigma.
\end{equation}

The bound holds for every graph, configuration, and system size. Thus a rule
cannot generate arbitrarily large attractive drift merely by producing more
neighbors, while endpoint symmetry preserves reciprocal pairwise drift.

The proof and an illustrative calculation for an isolated compact cluster
are given in the Supplemental Material \cite{SupplementalMaterial}.

\subsection{Three spatial replacement rules}

Each agent is replaced independently at rate \(\mu>0\), so its mean lifetime
is \(1/\mu\). Every event removes the departing agent and all its incident
edges, then inserts one new agent. If the old position was
\(\boldsymbol x\in\mathbb T_L^2\), the same position and bounded rules use a
displacement \(\boldsymbol\xi\in\mathbb R^2\) and set the new position to
\(\boldsymbol y=\boldsymbol x+\boldsymbol\xi\pmod L\). The global rule samples
\(\boldsymbol y\in\mathbb T_L^2\) directly. These operations induce
displacement kernels \(K\) on the torus. The three rules below differ only in
the choice of \(K\). After insertion, pairs inside the contact distance form new edges
immediately; no old edge is inherited.

For bounded relocation, let \(R\in(0,L/2)\) denote the largest allowed
displacement. The primary experiment uses \(R=5\sigma\); the radius scan
in Sec.~\ref{sec:radius_scan} varies \(R\) while keeping the other parameters fixed.

\begin{equation}
\label{eq:kernels}
\begin{aligned}
K_{\rm same}:&\quad \boldsymbol\xi=\boldsymbol0,\\
K_{\rm bounded}:&\quad \boldsymbol\xi\ \text{uniform in }
\{\boldsymbol z\in\mathbb R^2:|\boldsymbol z|\leq R\},\\
K_{\rm global}:&\quad \boldsymbol y\ \text{uniform on }\mathbb T_L^2.
\end{aligned}
\end{equation}

Here ``global'' describes the uniform destination distribution; replacement
events occur independently for each agent. The disk specifies the allowed
replacement displacements. All three rules
use the same event times and incident edge deletion procedure. The actual deleted edge sets can differ because the graphs evolve
differently. Replacement at the same position therefore isolates memory loss
without transport, while bounded and global relocation add different amounts
of mixing.

We use the dimensionless density \(\eta>0\), attraction strength
\(\mathrm{Pe}\geq0\), and mean lifetime \(\Theta>0\), defined by

\begin{equation}
\label{eq:parameters}
\eta=\frac{(N-1)\pi\sigma^2}{L^2},\qquad
\mathrm{Pe}=\frac{\chi\sigma^2}{\kappa},\qquad
\Theta=\frac{\kappa}{\mu\sigma^2}.
\end{equation}

Here \(\eta\) is the expected number of neighbors within one contact radius
in a spatially uniform population. The Péclet number \(\mathrm{Pe}\geq0\)
compares attraction with diffusion over that radius, and \(\Theta>0\) is the
mean agent lifetime in units of the diffusion time across one contact radius.
The simulations use \(L=1\) and \(\kappa=\sigma^2\). We have now specified
the motion, contact, and replacement rules introduced in
Fig.~\ref{fig:model}. The next section asks how their feedback treats broad and fine density
variations differently.

\section{How relocation and persistent contacts act on spatial scales}
\label{sec:theory}

\subsection{Relocation smooths different wavelengths differently}
\label{sec:transport}

We now turn the scale argument into a calculation. Any weak density pattern on
a periodic square can be decomposed into sinusoidal modes. For
\(\boldsymbol m=(m_x,m_y)\in\mathbb Z^2\), define the wavevector
\(\boldsymbol Q_{\boldsymbol m}=2\pi\boldsymbol m/L\), its magnitude
\(Q_{\boldsymbol m}=|\boldsymbol Q_{\boldsymbol m}|\), abbreviated as \(Q\)
for the selected mode, and the dimensionless
wavenumber \(u=Q_{\boldsymbol m}\sigma\geq0\). For
\(\boldsymbol m\ne\boldsymbol0\), the wavelength is
\(2\pi/Q_{\boldsymbol m}\). Small positive \(u\) means a broad pattern, while
large \(u\) means a fine pattern. The zero mode \(\boldsymbol m=\boldsymbol0\)
is the conserved mean density. A dimensionless growth rate
\(\lambda(u)\in\mathbb R\) is positive when structure at that scale is
amplified and negative when it is smoothed away.

Let
\(\rho:\mathbb T_L^2\times[0,\infty)\to[0,\infty)\) denote the ensemble
agent density, normalized by
\(\int_{\mathbb T_L^2}\rho(\boldsymbol x,t_{\rm phys})\,d^2\boldsymbol x=N\). A
replacement removes one agent at the departure position and inserts one after
a displacement drawn from the probability kernel \(K\). For
\(\boldsymbol x\in\mathbb T_L^2\), define the torus convolution
\((K*\rho)(\boldsymbol x)=\int_{\mathbb T_L^2}
\rho(\boldsymbol x-\boldsymbol\xi)K(d\boldsymbol\xi)\). Averaging this gain
and loss over many events gives the exact replacement contribution in physical
time,

\begin{equation}
\label{eq:renewal}
\left.\partial_{t_{\rm phys}}\rho\right|_{\rm ren}
=\mu(K*\rho-\rho).
\end{equation}

We now introduce the dimensionless time
\(t=\kappa t_{\rm phys}/\sigma^2\), and retain the symbol
\(\rho\) for the density expressed in this new time coordinate. We also write
\(X_i(t)=X_i(t_{\rm phys}=\sigma^2t/\kappa)\in\mathbb T_L^2\). Thus the simulation
time \(t\), its step
\(\Delta t>0\), and every growth rate reported below are dimensionless. A
physical rate is recovered by multiplication with \(\kappa/\sigma^2\). For a
density mode with scaled wavenumber \(u\), Eq.~\eqref{eq:renewal} becomes

\begin{equation}
\label{eq:renewal-rate}
\Lambda_{{\rm ren},K}(u)
=\Theta^{-1}[\widehat K(u)-1]
\end{equation}

where \(\Lambda_{{\rm ren},K}(u)\in\mathbb R\) is the contribution of
replacement to the dimensionless growth rate. For the isotropic kernels used
here, the Fourier multiplier is
\(\widehat K(\boldsymbol Q)=\int e^{-\mathrm i\boldsymbol Q\cdot
\boldsymbol\xi}K(d\boldsymbol\xi)\), where
\(\mathrm i=\sqrt{-1}\). It is real and depends only on \(u\), so we write
it as \(\widehat K(u)\). It is the
mean Fourier multiplier of a replacement displacement. Let
\(J_1:\mathbb R\to\mathbb R\) denote the Bessel function of the first kind
and order one, and let \(c_R=R/\sigma\in(0,L/(2\sigma))\) be the dimensionless
relocation radius. For the three rules,

\begin{equation}
\label{eq:kernel-transform}
\begin{aligned}
\widehat K_{\rm same}(u)&=1,\\
\widehat K_{\rm bounded}(u)&=\frac{2J_1(c_Ru)}{c_Ru},\\
\widehat K_{\rm global}(u)&=0\quad (u>0).
\end{aligned}
\end{equation}

The bounded expression has the continuous value
\(\widehat K_{\rm bounded}(0)=1\). Global relocation also has
\(\widehat K_{\rm global}(0)=1\), but its multiplier is zero for every
allowed nonzero mode on the torus. The uniform density is therefore conserved.
Long wavelengths correspond to small nonzero wavenumbers.

The three expressions encode three different physical operations. Replacement
at the same position does not move density and therefore does not directly
damp any wave. Bounded relocation acts locally: a wave much wider
than the relocation disk is only weakly affected, with a leading effect
proportional to \(u^2\): for \(QR=c_Ru\ll1\),
\(\Lambda_{{\rm ren},K}=-c_R^2u^2/(8\Theta)+O((c_Ru)^4)\).
The longest wave in the primary experiment has \(QR\simeq2.72\), outside
the regime of short displacements, so the predictions use the full disk
transform. Global relocation forgets the departure position and
adds the same decay rate to every nonuniform wave. Figure~\ref{fig:theory}(a) plots the
contributions from relocation alone after removing the Brownian term shared by
all three rules. The contrast at small \(u\) is the first theoretical reason
to expect replacement distance to control aggregation on large spatial scales.

A finite relocation range can produce stronger damping than global mixing.
The disk transform has negative lobes: at \(c_R=5\), its first negative
lobe occurs for \(0.766<u<1.403\). Replacements then transfer a negative mean Fourier
component, partly reversing the phase of the original density variation.
At these scales, the disk rule damps the mode more strongly than global
mixing, whose multiplier vanishes for every nonzero Fourier mode.

The simulator applies Brownian motion and a Bernoulli replacement update in
discrete steps \(\Delta t\). Let
\(p=1-e^{-\Delta t/\Theta}\in(0,1)\) be the replacement probability in one
step. At zero attraction, let \(\lambda_{0,K}^{\Delta t}(u)\in\mathbb R\)
denote the exact mean growth rate of the implemented update. When
\(1+p[\widehat K(u)-1]>0\), as for every condition used here, it is

\begin{equation}
\label{eq:null-rate}
\begin{aligned}
\lambda_{0,K}^{\Delta t}(u)
&=-u^2+\Delta t^{-1}
\log\left\{1+p[\widehat K(u)-1]\right\}.
\end{aligned}
\end{equation}

This rate calibrates the sampler and estimator before we measure the added
effect of attraction.

\subsection{How replacement shortens edge persistence}
\label{sec:persistence}

The replacement rule describes where density is moved, but attraction also
depends on how long an edge survives. We therefore ask a question for a single
pair before treating the full network: given the current separation of two
agents, what is the probability that their previous encounter still supports
an active edge?

At zero attraction, the relative separation of two independent Brownian agents
diffuses twice as fast as either agent alone. Replacement of either endpoint
deletes their old edge, so the edge is lost at total rate \(2\mu\). Between
the creation radius \(\sigma\) and deletion radius \(H\sigma\), these two
processes give a radial diffusion equation with a constant loss term. The
inner boundary has active probability one because a close encounter creates an
edge, and the outer boundary has probability zero because the edge is deleted.
Let \(\zeta=r/\sigma\in[0,\infty)\) denote dimensionless separation and let
\(h_\Theta(\zeta)\in[0,1]\) denote the conditional probability that the edge
is active in the stationary continuous time pair reference. In the annulus
\(1<\zeta<H\), this probability obeys

\begin{equation}
\label{eq:pair-profile}
h_\Theta''+\zeta^{-1}h_\Theta'-\Theta^{-1}h_\Theta=0,
\qquad h_\Theta(1)=1,\quad h_\Theta(H)=0.
\end{equation}

This boundary value problem has a unique solution between zero and one. We
complete it with \(h_\Theta(\zeta)=1\) for \(0\leq\zeta\leq1\) and
\(h_\Theta(\zeta)=0\) for \(\zeta\geq H\). Its closed form in modified Bessel
functions is given in the Supplemental Material \cite{SupplementalMaterial}. Here
primes denote derivatives with respect to \(\zeta\). The
solution is exact for one independent Brownian pair at zero attraction in a
homogeneous background, with exponential replacement and immediate edge
activation. The three replacement rules have the same pair profile in that
reference state because all delete an incident edge at the same rate and all
preserve a homogeneous density. Correlations generated by attraction are not
included at this stage.

\subsection{From edge persistence to collective attraction}

The full graph contains many correlated binary edge variables. To obtain a
tractable density equation, we replace the instantaneous graph by the average
probability that a pair is connected at a given separation. In this annealed mean field approximation, the density pattern responds to
the averaged contact network. The
resulting radial force weight is
\(\psi_\Theta(\zeta)=h_\Theta(\zeta)\widetilde q(\zeta)\), where
\(\widetilde q(\zeta)=q(\sigma\zeta)\). Both functions map
\([0,\infty)\) to \([0,1]\). Because the force weight depends only on
separation, its Fourier transform in two dimensions reduces to a Hankel
transform in one dimension, the radial analogue of a Fourier transform. Let
\(J_0:\mathbb R\to\mathbb R\) denote the Bessel function of the first kind
and order zero. We define

\begin{equation}
\label{eq:hankel}
F_\Theta(u)=2\pi\int_0^H \zeta\psi_\Theta(\zeta)J_0(u\zeta)\,d\zeta.
\end{equation}

Thus \(F_\Theta:[0,\infty)\to\mathbb R\), and
\(F_\Theta'(u)=dF_\Theta/du\).

Let \(a=\eta/\pi>0\), let \(\bar s=aF_\Theta(0)\geq0\) be the mean weighted
degree in the homogeneous closure, and let \(d_0=s_0+2\bar s>0\). We now
perturb the uniform density by one weak wave and
retain only terms proportional to its amplitude. This linearization must
include both the change in the number of attractive neighbors and the change
in the normalization denominator. It gives the response per unit attraction
strength, \(G_\Theta:[0,\infty)\to\mathbb R\), so that the attraction
contribution is \(\mathrm{Pe}\,G_\Theta(u)\):

\begin{equation}
\label{eq:attraction}
G_\Theta(u)=
a\,u[-F_\Theta'(u)]
\left[
d_0^{-1}-aF_\Theta(u)d_0^{-2}
\right].
\end{equation}

The first bracketed term is the direct increase of attractive flux in a denser
region. The second subtracts the part offset by the strength normalization:
the same density increase also enlarges the denominator in Eq.~\eqref{eq:motion}. This
second term is why the normalization must be carried consistently into the
collective theory. The closed density equation and full algebra are given in
the Supplemental Material; the main text keeps the resulting response and its
physical interpretation. Radial symmetry gives \(F_\Theta'(0)=0\), so
\(G_\Theta(0)=0\) and the conserved mean density is not altered.

\begin{figure*}[t]
  \centering
  \includegraphics[width=\textwidth]{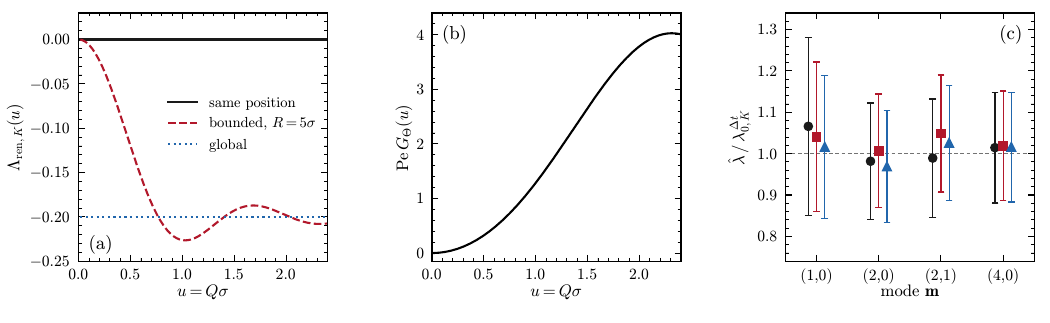}
  \caption{\textbf{Relocation and persistent contacts make separate contributions to the growth of density modes.}
  (a) Replacement contribution after removing the Brownian term, plotted as
  $\lambda_{0,K}^{\Delta t}(u)+u^2$. This is the discrete update counterpart
  of $\Lambda_{{\rm ren},K}(u)$ in Eq.~\eqref{eq:renewal-rate}, with $\Delta t=10^{-3}$,
  $\Theta=5$, and $R=5\sigma$ for the disk rule.
  Negative values mean damping. Same position replacement leaves density
  unchanged, bounded relocation damps different wavelengths unequally, and
  global relocation adds a decay rate $\Theta^{-1}$ to each nonzero mode.
  The legend also identifies the symbols in panel (c).
  (b) Attraction contribution $\mathrm{Pe}\,G_\Theta(u)$ at
  $\mathrm{Pe}=10.5$, $\eta=6$, $H=2$, and $s_0=1$, predicted by the mean field closure from the pair
  activity probability in Eq.~\eqref{eq:pair-profile}. Adding this term to the exact baseline
  gives Eq.~\eqref{eq:dispersion}.
  (c) Calibration without attraction: measured rate divided by the exact
  rate of the discrete update [Eq.~\eqref{eq:null-rate}] for the three rules and four modes.
  The dashed line at one denotes exact agreement. Error bars are simultaneous
  95\% bootstrap intervals; all twelve contain one. Increasing $u$
  corresponds to a shorter wavelength.}
  \label{fig:theory}
\end{figure*}

Figure~\ref{fig:theory}(b) shows how this amplification rises over the long
and intermediate scales tested here.

Combining the continuous time attraction approximation with the exact
discrete time baseline gives the tested approximation for
the total dimensionless growth rate \(\Lambda_K^{\Delta t}(u)\in\mathbb R\)
as

\begin{equation}
\label{eq:dispersion}
\Lambda_K^{\Delta t}(u)
\simeq \lambda_{0,K}^{\Delta t}(u)
+\mathrm{Pe}\,G_\Theta(u).
\end{equation}

Equation~\eqref{eq:dispersion} is the central prediction. The exact first term combines
Brownian smoothing and replacement transport. The approximate second term is
the amplification produced by the contact network. Using this continuous
time response at finite \(\Delta t\) introduces an additional discretization
approximation. A finite range of growing
spatial scales requires their sum to be negative for broad patterns, positive
at an intermediate scale, and negative again for fine patterns, where
diffusion dominates.

Figure~\ref{fig:theory}(c) calibrates the baseline without attraction, with
unity denoting agreement between measured and exact rates. The attraction
term neglects the response time of the edge network, changes in pair
structure, and fluctuations of the normalization denominator. We test it
separately before judging the sum near a sign change.

\section{Measuring growth at selected spatial scales}
\label{sec:design}

\subsection{Design}

We test individual density waves by imposing one weak sinusoidal modulation
and measuring whether its contrast grows or decays. The primary experiment
used \(N=256\), \(\eta=6\), \(H=2\), \(s_0=1\), \(\Theta=5\), and
\(\mathrm{Pe}\in\{0,10.5\}\), with \(\Delta t=10^{-3}\), final time
\(t=0.6\), and records every 0.01 time unit. These parameters give
\(\sigma=\sqrt{6/(255\pi)}\simeq0.0865\) and
\(R=5\sigma\simeq0.433<L/2\). The prescribed modes
\(\boldsymbol m\in\{(1,0),(2,0),(2,1),(4,0)\}\) correspond to
\(u\simeq\{0.544,1.088,1.216,2.175\}\). The three primary modes,
\((1,0)\), \((2,0)\), and \((4,0)\), have horizontal wavelengths
\(L\), \(L/2\), and \(L/4\). Decay, growth, and decay across these scales
under displaced replacement would support a finite range of initial
amplification. Replacement at the same position controls for memory loss
without transport. The nearby mode \((2,1)\) was selected in advance as a secondary probe of
the intermediate response.

Initial positions were sampled independently from
\(\rho_{\boldsymbol m}(\boldsymbol x,0)/N\), using rejection sampling, with
initial modulation depth \(\epsilon\in(0,1)\):

\begin{equation}
\label{eq:initial-density}
\rho_{\boldsymbol m}(\boldsymbol x,0)=\frac{N}{L^2}
\left[1+\epsilon\cos(\boldsymbol Q_{\boldsymbol m}\cdot\boldsymbol x)\right],
\qquad \epsilon=0.08\in(0,1).
\end{equation}

The amplitude of the imposed wave is

\begin{equation}
\label{eq:amplitude}
A_{\boldsymbol m}(t)=2\operatorname{Re}\left[
\frac{1}{N}\sum_{j=1}^N
e^{-\mathrm{i}\boldsymbol Q_{\boldsymbol m}\cdot X_j(t)}
\right].
\end{equation}

Here \(\operatorname{Re}\) denotes the real part and
\(A_{\boldsymbol m}(t)\in[-2,2]\) is dimensionless. It is the signed
projection onto the imposed phase; fitting its ensemble mean measures the
early evolution of density contrast at that scale.
Initial edges were sampled from Eq.~\eqref{eq:pair-profile}, the reference
profile without attraction, rather than from a stationary interacting state.
The independent random streams and update order are specified in
Appendix~\ref{sec:methods}.

Each of the 3,200 independent replicate blocks contained all three rules,
four modes, and two attraction conditions, giving 76,800 trajectories.
Prescribed random number streams were shared within blocks; initial
positions were bitwise identical across rules and attraction conditions for
each mode. The preliminary study used different seeds and supplied no
trajectories to these estimates. It fixed the estimator, sample size, and
attraction strength before the present outcomes were available. The chosen
\(\mathrm{Pe}=10.5\) lies within the theoretical feasible interval
\(9.791<\mathrm{Pe}<11.804\) for the prescribed signs, halfway between two
values used in that preliminary study.

\subsection{Growth rate estimators}

Let \(A_\ell^P(t)\) and \(A_\ell^0(t)\), with
\(\ell\in\{1,\ldots,3200\}\) in the primary experiment, denote amplitudes aligned with the imposed
phase in paired trajectories at \(\mathrm{Pe}=10.5\) and \(\mathrm{Pe}=0\),
respectively; \(P\) labels the condition with attraction. For each condition,
the unadjusted estimator fits its ensemble mean \(\overline A(t)\) to

\begin{equation}
\label{eq:exp-fit}
\overline A(t)=c e^{\lambda t},\qquad c\geq0,
\end{equation}

where \(c\in[0,\infty)\) is the fitted initial amplitude and
\(\lambda\in\mathbb R\) is the dimensionless growth rate. The prespecified
fit window depends on \(u\), not on observed outcomes, and is defined in
the Supplemental Material \cite{SupplementalMaterial}.

The change caused by attraction is smaller than the fluctuations among
individual trajectories. We reduce this noise without changing the target
quantity by pairing every attraction run with a run at zero attraction that
starts from the same configuration and uses matched random number streams.
We subtract their shared fluctuation and add back the exact mean at zero
attraction. In statistical terms this is a control variate whose coefficient
is fixed to one. For each replicate and recorded time \(t=n\Delta t\), with
\(n\in\mathbb N_0=\{0,1,2,\ldots\}\),

\begin{equation}
\label{eq:cv}
Z_\ell(t)=A_\ell^P(t)-
\left[A_\ell^0(t)-A_\ell^0(0)e^{\lambda_{0,K}^{\Delta t}(u)t}\right],
\end{equation}

where \(Z_\ell(t)\in\mathbb R\) is the adjusted amplitude. Its ensemble
mean \(\overline Z(t)\) is fitted with the same model and
window as Eq.~\eqref{eq:exp-fit}. Writing \(\mathbb E\) for expectation over
replicate trajectories, the subtracted fluctuation has zero mean, so
\(\mathbb E[Z_\ell(t)]=\mathbb E[A_\ell^P(t)]\). The adjusted estimator
therefore targets the same mean growth curve as the unadjusted fit. It uses
the exact baseline at zero attraction verified below; it does not insert the
attraction approximation from Eq.~\eqref{eq:dispersion}. We report both
estimates whenever they lead to different inferential conclusions.

For a separate quantitative test of Eq.~\eqref{eq:attraction}, we fit an
attraction increment directly to the paired amplitude difference between
trajectories with and without attraction, holding their measured initial
amplitude fixed. This estimator was specified in the preliminary study.
For curves that depart from a single exponential, it differs from subtracting
the exact baseline rate from the free amplitude total rate above.
We compare the paired increment with \(\mathrm{Pe}\,G_\Theta(u)\), evaluated without
fitting, in the six combinations of replacement rule and mode at the long
and primary intermediate scales. The summary measures are the mean absolute
percentage error and six pairwise differences among replacement rules,
normalized by the common theoretical response. The latter are compared with
the prespecified equivalence range \([-0.15,0.15]\). Full definitions are in the
Supplemental Material \cite{SupplementalMaterial}.

\subsection{Simultaneous uncertainty and the joint criterion}

The independent replicate block is the statistical unit. We resample complete matched blocks 10,000 times,
recomputing every nonlinear fit. Each statistic and joint family must remain
computable in at least 99\% of draws. Simultaneous intervals use the largest
standardized bootstrap deviation within each family, protecting all its
comparisons together. The primary family contains nine rates: three rules
at the long, intermediate, and short modes, with the required signs in
Table~\ref{tab:signs}.

\begin{table}[t]
\caption{Prespecified signs of the primary density responses. A plus sign
denotes growth and a minus sign denotes decay. All nine simultaneous
intervals must resolve their assigned signs for the joint criterion to pass.}
\label{tab:signs}
\begin{ruledtabular}
\begin{tabular}{lccc}
Mode & Same position & Disk & Global\\
\hline
Long, $(1,0)$ & $+$ & $-$ & $-$\\
Intermediate, $(2,0)$ & $+$ & $+$ & $+$\\
Short, $(4,0)$ & $-$ & $-$ & $-$
\end{tabular}
\end{ruledtabular}
\end{table}

Every simultaneous 95\% interval must resolve its assigned sign. The joint
criterion also requires recovery of the exact zero attraction baselines,
positive contrasts at the longest wavelength, and the specified attraction
accuracy and equivalence among rules. All requirements must hold jointly.
Statistical details are in Appendix~\ref{sec:methods} and the Supplemental
Material \cite{SupplementalMaterial}.

\subsection{Additional experiments}
\label{sec:supplemental-design}

A separate study tests eight disk radii between \(c_R=1\) and \(5.5\),
alongside replacement at the same position and global relocation. It uses
the longest mode, both attraction conditions, and 800 new independent matched
blocks. Its ten rates with attraction and nine contrasts with replacement
at the same position form separate simultaneous 95\% families. Every disk
satisfies \(R<L/2\). The scan resolves signs at the sampled radii.

Another 400 independent blocks compare the original settings with
\(\Delta t/2\) and \(\epsilon/2\), for all three rules at
\(\boldsymbol m=(1,0)\) and the disk rule at \((2,0)\). The eight paired rate changes form one simultaneous 95\% family measuring
sensitivity to the time step and modulation amplitude.

To sample the scale dependence more closely, a further independent design
adds modes \((1,1)\), \((2,2)\), \((3,0)\), \((3,1)\), and \((3,2)\) under all three rules,
with 3,200 new matched blocks at \(\mathrm{Pe}=0\) and \(10.5\).
It retains the original initialization, time step, fit windows, and
estimators. Its fifteen rates with attraction, fifteen zero attraction
baselines, and fifteen attraction response ratios form three separate
simultaneous 95\% families. A distinct set of 1,600 blocks compares
\(\epsilon=0.08\) with \(0.16\) for the three rules at \((1,0)\)
and the disk rule at \((2,0)\) and \((2,2)\). Its five paired rate
differences form another simultaneous family. All additional designs fix
their sample sizes, seeds, and analysis rules before production; none changes
the original joint criterion. Implementation, the interpretation of amplitude
dependence, and local structure measurements are described in the Supplemental Material
\cite{SupplementalMaterial}.

A retrospective reanalysis compares the earliest recorded window,
$0\le t\le0.1$, with the original fit windows. Its specification was fixed
after earlier summaries from all three cohorts had been examined. Each
cohort retains separate blocks and two simultaneous 95\% families, one for
early rates and one for paired changes between windows. A further family
compares attraction responses among rules at the five added modes. The
Supplemental Material gives the full retrospective specification and
confidence families.

\section{Results}
\label{sec:results}

\begin{figure*}[t]
  \centering
  \includegraphics[width=\textwidth]{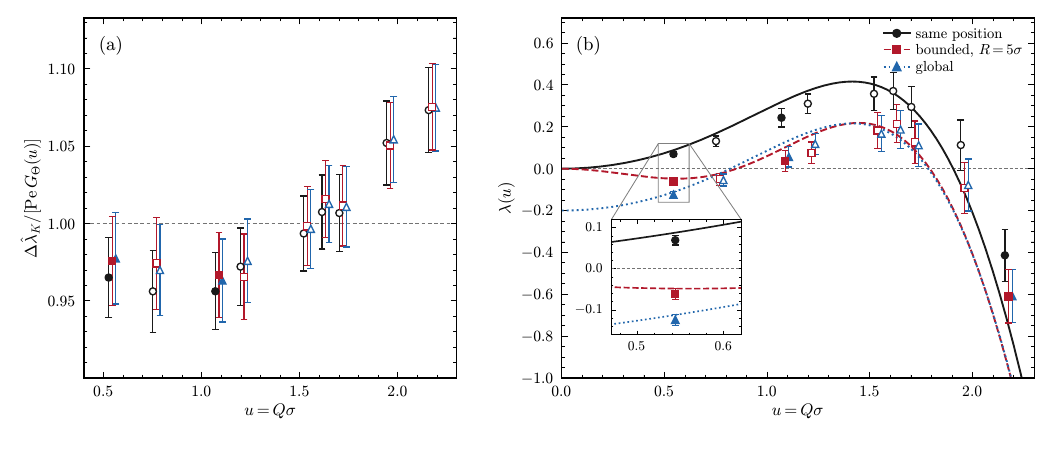}
  \caption{\textbf{Replacement controls the scale dependence of the early response.}
  (a) Measured growth added by attraction divided by its unfitted prediction,
  across all nine sampled wavenumbers. Unity denotes agreement.
  (b) Total fitted rate versus scaled wavenumber. Curves show
  Eq.~\eqref{eq:dispersion} at $\mathrm{Pe}=10.5$ without fitting; the global
  curve applies to nonzero modes only. Circles, squares, and triangles denote
  replacement at the same position, within the disk, and globally in both
  panels. Filled symbols identify the six primary response comparisons in
  (a) and the nine primary sign comparisons in (b); open symbols identify
  conditions outside those criteria. Horizontal offsets separate rules at
  the same $u$ only; the inset magnifies $(1,0)$ without offsets.
  Each panel contains 27 estimates. Error bars are simultaneous 95\%
  intervals within separate families: the twelve original responses and
  fifteen new responses in (a); the nine primary, three secondary, and
  fifteen new rates in (b). The original and new cohorts each contain
  3,200 independent paired blocks with 10,000 whole-block resamples;
  intervals are not jointly simultaneous across cohorts. Comparison of
  these rates over the full windows with the earliest rates is shown in
  Fig.~\ref{fig:initial_window}.}
  \label{fig:growth}
\end{figure*}

We test the exact transport baseline before comparing the added attraction
response and the resulting total rates.

\subsection{Dynamics at zero attraction recover the exact replacement baseline}
\label{sec:baseline}

Equation~\eqref{eq:dispersion} combines an exact baseline with an approximate attraction term.
We therefore test the baseline first. At \(\mathrm{Pe}=0\), all twelve
simultaneous 95\% intervals contain the exact rates for the update implemented in discrete time from Eq.~\eqref{eq:null-rate}
for the three replacement rules and four tested modes [Fig.~\ref{fig:theory}(c)]. This agreement
checks the displacement samplers, the discrete replacement update, and the
growth rate estimator before any network force is introduced. All fifteen
additional baseline intervals also contain their exact rates
(Supplemental Material \cite{SupplementalMaterial}).

\subsection{Attraction contributes a similar growth increment under all three replacement rules}

The second test isolates the approximate part of Eq.~\eqref{eq:dispersion}. Over the six
prespecified combinations of replacement rule and long or intermediate mode,
the measured growth added by attraction differs from the theoretical
prediction by 3.26\% on average. The bootstrap 95\% upper bound
on this mean error is 4.38\%, below the prespecified bounds of 10\% for the point
estimate and 15\% for its upper limit [Fig.~\ref{fig:growth}(a)].
All six measured responses are slightly below the corresponding prediction.

We also compare the measured attraction response directly among replacement
rules. All six pairwise normalized differences have simultaneous 90\%
intervals inside the prespecified \([-0.15,0.15]\) equivalence range; the most
extreme interval endpoints are approximately \(-0.025\) and \(0.013\).
Changing where a replacement appears shifts the exact baseline, while the
added growth varies only weakly within the tested resolution. This is the
empirical basis for treating the two contributions as approximately
separable.

The supplementary comparison at the five added modes gives fifteen
normalized differences, all with simultaneous 90\% intervals inside the
same range; their extreme endpoints are $-0.033$ and $0.018$.
This analysis was frozen retrospectively and inherits the original
equivalence margin, which was not part of the additional mode protocol.
The two remaining original modes also lie within that range in the
existing simultaneous 95\% family of twelve contrasts. Together, these separate comparisons support a similar attraction
response among rules at all nine sampled wavelengths. Small
offsets are tabulated in the Supplemental Material.

\subsection{Replacement distance reverses the longest tested wave}

\begin{figure}[!tbp]
\centering
\includegraphics[width=\columnwidth]{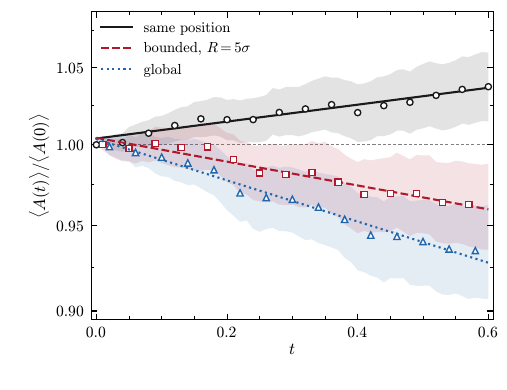}
\caption{\textbf{Raw evolution of the longest density pattern.}
The horizontal axis is dimensionless time; the logarithmic vertical axis
is the raw ensemble projected amplitude divided by its initial ensemble
mean, for $\boldsymbol m=(1,0)$ and $\mathrm{Pe}=10.5$.
Symbols sample recorded means at distinct recorded times for each rule to
avoid overlap; markers remain at their actual recording times. Lines are free amplitude
exponential fits using every $0.01$ record over $0\le t\le0.6$.
Shading gives pointwise 95\% intervals from 10,000 resamples of the
3,200 original paired blocks, not simultaneous bands. Neither the data nor
the fitted lines use the control variate. The supplementary two-panel
version also shows the zero attraction controls.}
\label{fig:raw_longwave}
\end{figure}

We next examine their sum. The longest tested wave, \(\boldsymbol m=(1,0)\) with
\(u\simeq0.544\), spans the system in one direction. It grows under
replacement at the same position but decays when replacement is displaced
(Table~\ref{tab:longwave}). Figure~\ref{fig:raw_longwave} shows the underlying
signal directly: the mean density contrast increases for replacement at the
departure site and decreases for both displaced rules.

\begin{table}[!tbp]
\caption{Early response of the longest tested density wave,
$\boldsymbol m=(1,0)$, in the original experiment. Rates are control variate
estimates; intervals belong to the nine comparison simultaneous 95\% family.}
\label{tab:longwave}
\begin{ruledtabular}
\begin{tabular}{lrc}
Replacement rule & $\widehat\lambda$ & 95\% interval\\
\hline
Same position & $0.070$ & $[0.058,0.082]$\\
Disk, $R=5\sigma$ & $-0.061$ & $[-0.074,-0.047]$\\
Global & $-0.124$ & $[-0.138,-0.110]$
\end{tabular}
\end{ruledtabular}
\end{table}

Here \(\widehat\lambda\in\mathbb R\) denotes the estimated dimensionless
growth rate. The paired differences are
\(0.131\,[0.124,0.137]\) between replacement at the same position and bounded
relocation and \(0.194\,[0.187,0.200]\) between replacement at the same
position and global relocation. With the edge deletion procedure held fixed, relocating replacements
reverses the sign at the longest wavelength. Global relocation gives the
strongest damping [Fig.~\ref{fig:growth}(b)]. Equation~\eqref{eq:renewal-rate} predicts the ordering
among replacement rules, while Eq.~\eqref{eq:dispersion} predicts the total signs.

The unadjusted point estimates have the expected direction in all nine primary
conditions, but their simultaneous intervals do not resolve the long mode
under replacement at the same position or two intermediate modes. The sign
reversal at the longest wavelength above is
therefore inferred with the prespecified paired variance reduction. Its calibration uses the exact zero attraction baseline verified in
Sec.~\ref{sec:baseline}.

\subsection{Intermediate and short waves}
\label{sec:intermediate}

Suppressing the longest wave does not by itself select a finite scale.
We must also test for growth at intermediate wavelengths and decay at
shorter wavelengths. At
\(\boldsymbol m=(2,0)\), replacement at the same position and global
relocation have resolved positive rates,
\(0.243\,[0.199,0.288]\) and
\(0.057\,[0.008,0.105]\). Bounded relocation has a positive point
estimate, \(\widehat\lambda_{\rm bounded}=0.038\), but its interval
\([-0.012,0.088]\) includes zero.

At the nearby secondary mode, \(\boldsymbol m=(2,1)\), the control variate
intervals are positive under all three rules.

At the short mode \(\boldsymbol m=(4,0)\), all three rules give decay. The estimates are
\(-0.414\), \(-0.610\), and \(-0.608\), and the largest simultaneous upper
limit is \(-0.289\). This resolves decay at the shortest tested wavelength,
consistent with Brownian smoothing [Fig.~\ref{fig:growth}(b)]. The primary
experiment resolves eight of its nine prescribed signs, so its joint
criterion remains unmet.

The independent additional modes now resolve growth under both displaced
rules at \(u=1.538,1.631,1.720\), and decay at \(u=0.769\).
For example, at \(u=1.538\) the bounded rate is
\(0.182\,[0.097,0.267]\), and the global rate is
\(0.168\,[0.082,0.253]\). At \(u=1.961\), however, all three
intervals include zero. Together with the original data, the sampled signs
place the lower change between \(0.769\) and \(1.216\) for bounded
relocation, and between \(0.769\) and \(1.088\) for global relocation.
For both, the upper change lies between the resolved growth at \(1.720\)
and decay at \(2.175\). These brackets combine resolved signs from separate confidence families.
They locate intermediate amplification between sampled damped long and
short scales; resolving a continuous unstable band requires denser sampling
and a joint analysis of its boundaries. The corresponding closure zeros, approximately \(0.85,1.79\)
for bounded and \(0.82,1.78\) for global relocation, lie inside these
brackets. Replacement at the same position remains positive at every
resolved sampled point below \(1.961\), with its upper change likewise
between \(1.720\) and \(2.175\).

\subsection{Early agreement and changes with the observation window}
\label{sec:cancellation}

\begin{figure}[t]
\centering
\includegraphics[width=\columnwidth]{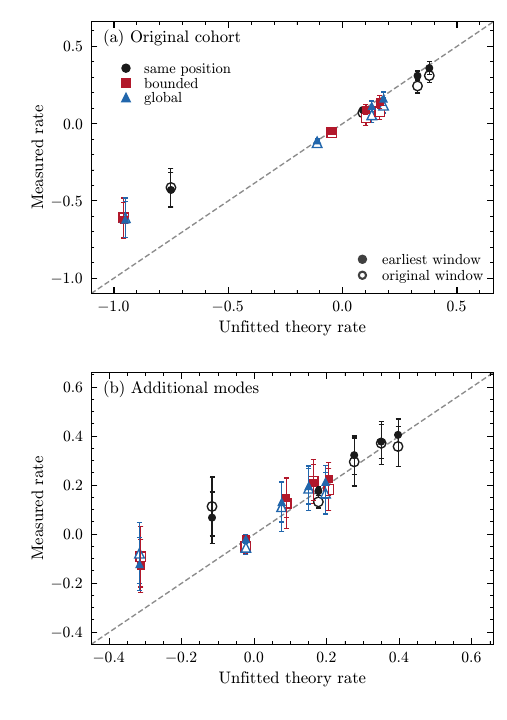}
\caption{\textbf{Early and full observation windows give different rate estimates.}
Measured control variate rates are plotted against the unfitted
Eq.~\eqref{eq:dispersion}: (a) twelve original and (b) fifteen additional
conditions, with different axis ranges. Filled symbols use $0\le t\le0.1$;
open symbols use the original mode dependent windows. Gray segments pair
conditions; the diagonal denotes equality. Circles, squares, and triangles
denote same position, disk, and global replacement. Bars are simultaneous
95\% intervals: separate early families of twelve and fifteen rates, and
full-window families of nine primary plus three secondary rates in (a)
and fifteen in (b). Each cohort has 3,200 blocks. These families are not
pooled; paired window-change intervals are in the Supplemental Material.}
\label{fig:initial_window}
\end{figure}

How much of the discrepancy with Eq.~\eqref{eq:dispersion} is already
present at the start, and how much accompanies a longer observation
window? Over the original full windows, only two of nine primary
predictions lie inside their simultaneous intervals. Figure~\ref{fig:initial_window}
compares those fits with rates obtained from the first eleven records,
$0\le t\le0.1$, using the same free amplitude control variate estimator.
These effective rates summarize the response over the earliest observed
interval. The full-window discrepancy is the sum of the early discrepancy
and the paired change between windows, giving a descriptive comparison
of the two fits.

At long and intermediate scales, the early simultaneous intervals contain
the theory in all nine original conditions and all twelve additional
conditions with $u\le1.720$. All ten radius conditions also contain the
theory. The largest absolute early point discrepancies are $0.022$,
$0.061$, and $0.007$, respectively. The largest absolute discrepancies
relative to the theoretical attraction contribution are about $1.2\%$,
$1.9\%$, and $1.9\%$, respectively.
Thus the earliest rates are compatible with the closure across these
scales, with different precision in the three cohorts.

The paired changes reveal a less uniform pattern. All nine original
long and intermediate point estimates decrease when the full window is
used, but only five changes have simultaneous intervals below zero.
Among the twelve corresponding additional conditions, eleven point
estimates decrease and only three decreases are resolved, all at
$u=0.769$. The exception in direction is the disk rule at $(3,0)$,
whose change is $0.004\,[-0.070,0.078]$. All ten radius estimates
decrease, but none of those paired intervals excludes zero. Downward changes are common in the point estimates and statistically
resolved in a subset. For several additional intermediate conditions, both
window estimates lie above theory, and the decrease brings the full-window
estimate closer to the prediction.

The shorter scales show a different discrepancy. At $u=1.961$ and
$2.175$, all six early intervals lie above the closure prediction.
The early differences are $0.183$ to $0.194$ and $0.321$ to $0.330$,
respectively, or about $4.9$--$5.2\%$ and $8.1$--$8.3\%$ of the
predicted attraction contribution. The mismatch is therefore already
present within the earliest observed window. The closure still averages over random endpoint strengths and network
correlations, even with initial edges sampled from the reference profile.

Cancellation makes these discrepancies important for the total rate.
Diffusion and relocation supply a negative contribution and attraction
a positive contribution of similar magnitude; the total is their small
remainder. A small fractional discrepancy in attraction can substantially
shift that remainder near a sign change. The window comparison distinguishes
an early residual from a later change, but does not decide whether edge
relaxation, pair correlations, or normalization fluctuations produce them.
The structural diagnostics below examine what changes during these windows.

\subsection{Varying the relocation radius changes growth into decay}
\label{sec:radius_scan}

\begin{figure*}[t]
  \centering
  \includegraphics[width=\textwidth]{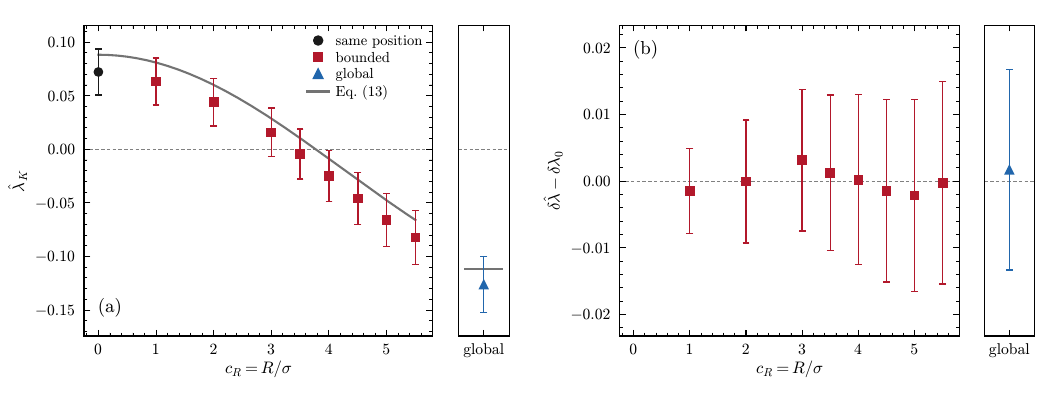}
  \caption{\textbf{The relocation radius controls the longest wavelength response.}
  (a) Fitted rate at $\boldsymbol m=(1,0)$ and $\mathrm{Pe}=10.5$ versus
  $c_R=R/\sigma$. The same position control is shown at zero; global
  relocation is a separate categorical reference, not a numerical radius.
  The gray curve is the unfitted closure in Eq.~\eqref{eq:dispersion}.
  (b) Measured rate contrast minus its exact baseline contrast, where
  $\delta\widehat\lambda=\widehat\lambda_K-\widehat\lambda_{\rm same}$
  and $\delta\lambda_0=\lambda_{0,K}^{\Delta t}-\lambda_{0,\rm same}^{\Delta t}$.
  Zero means agreement with the change predicted by transport alone.
  Rates are fitted using the control variate in Eq.~\eqref{eq:cv}.
  Error bars are simultaneous 95\% intervals across the ten rates in (a)
  and, separately, the nine paired contrasts in (b), from 800 independent
  blocks and 10,000 bootstrap resamples.}
  \label{fig:radius}
\end{figure*}

The independent radius scan tests whether the reversal extends beyond the
three original rules. Figure~\ref{fig:radius}(a) shows a decreasing response
as the disk radius increases. The longest mode has resolved positive rates
at \(R/\sigma=1\) and \(2\), and resolved negative rates at
\(4.5,5,5.5\). At \(4\), the interval is also negative but its upper
limit is only \(-0.001\). The intervals at \(3\) and \(3.5\) include zero.
The controls reproduce growth at the same position and decay under global
relocation. Thus radius controls the sign over this sampled range, but the
measurements do not identify a precise critical radius.

All nine paired contrasts with replacement at the same position are
negative. To isolate their origin, panel (b) subtracts the exact change in
the baseline from each measured contrast. The residual estimates range from
\(-0.0022\) to \(0.0031\), and all nine simultaneous intervals contain
zero. These residuals are compatible with replacement transport accounting for
the radius dependence while the attraction response remains approximately
shared.
The gray curve in panel (a) uses Eq.~\eqref{eq:dispersion} without fitting;
the numerical rates and their unadjusted counterparts are given in the
Supplemental Material \cite{SupplementalMaterial}.

\subsection{Robustness of the early response interpretation}
\label{sec:robustness}
\label{sec:window}

The reversal of the longest wave persists across four retrospective fit
windows, each evaluated with a separate simultaneous 95\% family of twelve
conditions. Some intermediate signs lose resolution in later windows,
consistent with the window dependence above. The complete rate comparisons
and descriptive fits over overlapping windows are given in the Supplemental
Material \cite{SupplementalMaterial}.

Local structure also changes during these fits. In the independent
radius study at the longest mode, the number of pairs within one
contact radius at \(t=0.6\) is about 45\% above the independent uniform
reference under replacement at the same position, and about 40\% above it
under both displaced rules. Without attraction, these ratios remain close
to one. The mean active degree also increases with attraction under every
rule. Local aggregation therefore develops even while the broad density pattern
decays. Its role in the rate discrepancy is considered in the Discussion.

The 400-block time step check gives changes below \(0.0004\) for the
longest mode, with simultaneous intervals within \([-0.0025,0.0032]\);
the disk \((2,0)\) change is \(-0.0014\,[-0.0112,0.0084]\).
In the new 1,600-block amplitude test, doubling \(\epsilon\) from
\(0.08\) to \(0.16\) gives five rate differences whose simultaneous
intervals all include zero. The longest-mode differences are about
\(0.01\), with intervals contained in \([-0.011,0.031]\); the two
intermediate comparisons are less precise. Assuming a leading quadratic
amplitude correction with negligible higher orders, dividing these
differences and intervals by three estimates the correction at \(0.08\).
The largest absolute interval endpoint is then below \(0.0103\) for the
longest mode, and below \(0.0334\) and \(0.0406\) for the two disk modes,
respectively.
These estimates quantify the leading correction under the stated
assumptions; they are not strict bias bounds. The complete amplitude
comparisons and expansion assumptions are given in the Supplemental
Material \cite{SupplementalMaterial}.

\section{Discussion}
\label{sec:discussion}

Replacement acts on a density pattern through the distribution of the
successor's displacement. Its Fourier transform determines the transfer of
each modal density component, while the replacement rate sets how often
this transport occurs. Persistent contacts provide a competing attraction
response. Across nine sampled wavelengths, the separate original and
supplementary comparisons find only small differences in that response
among replacement rules. This
approximate separation explains why changing replacement locations can
reverse the early evolution of a pattern spanning the system even when the
replacement rate and edge deletion procedure are unchanged.
The measurements test the full displacement transform, rather than the
small-disk diffusion approximation described in Sec.~\ref{sec:transport}.

The prediction concerns a prepared response. The weak initial density
modulation and zero attraction edge profile specify an initial condition
for the interacting dynamics. The fitted rates characterize its early
relaxation, not a stationary interacting background. The earliest measured rates at long and
intermediate scales are compatible with the closure within the available
precision, while both short modes already show discrepancies. Longer windows
lower many point estimates, with resolved changes in only a subset. Contacts
accumulate during the same period, but distinguishing delayed edge response
from pair correlations and strength fluctuations requires further analysis.
The model of point agents also omits excluded volume, limiting direct comparison
with crowded physical particles.

Suppression of broad patterns, amplification at intermediate scales, and
smoothing at short scales provide a possible route to wavelength selection
at the initial growth stage
\cite{CrossHohenberg1993PatternFormation,Glotzer1995Reaction,Cates2010Arrested}.
The eventual domain size also depends on nonlinear coarsening and its
arrest, as studies of active and reaction diffusion systems demonstrate
\cite{Tjhung2018Cluster,Brauns2021Wavelength}.
The new independent modes resolve intermediate growth under both displaced
rules and place it between sampled long and short scales with decay.
The discrete sign brackets identify the sampled scales over which the
response changes sign.
Separate longer simulations at a slightly higher attraction strength under global
relocation show clustering at a nonzero characteristic wavenumber, with
similar spectral statistics across the tested sizes, but retain slow
fluctuations on the largest scales \cite{SupplementalMaterial}.
Whether this clustering reaches a stationary pattern or retains a fixed
length scale remains open.

Evolving the mean edge state together with density offers a route to
including contact delays and changes in local pair structure. Substituting
measured radial moments into the present static closure gives no consistent
improvement (Supplemental Material \cite{SupplementalMaterial}), suggesting
that this correction alone misses part of the response. A dynamical closure
can be tested against the measured changes between observation windows.
Its extension to stationary interacting backgrounds, and the dependence on
density, lifetime and population size, remain open.

\raggedbottom
\section{Conclusion}
\label{sec:conclusion}

Replacement both removes old contacts and transports population to new
positions. We have shown that changing the second effect can reverse the
early evolution of a broad density pattern while keeping replacement
frequency and edge removal fixed. The broadest tested pattern grows when
successors remain at the departure position and fades under the two
displaced rules; the radius scan traces this change from growth to decay.
Replacement frequency alone therefore does not determine which density
contrasts are initially amplified.

Our description combines exact diffusion and replacement transport with an
approximate attraction term. Paired tests support similar attraction
increments across the three rules at nine sampled wavelengths, while the
radius scan independently supports the transport mechanism. Rates over the earliest finite window are compatible with the
averaged theory at long and intermediate scales within the available
precision; discrepancies at short scales show where the approximation
needs refinement.
Cancellation between attraction and smoothing makes these errors important
for a small total rate.

The data resolve intermediate amplification between sampled damped scales
under displaced replacement. The original joint criterion remains unmet,
with eight of nine prescribed signs resolved. These results describe early
density responses; their persistence at long times remains open. A theory
that evolves contact activity together with density can connect the observed
changes in early rates to the later nonlinear organization.

\appendix
\section{Methods and reproducibility details}
\label{sec:methods}

\subsection{Numerical integration and event ordering}

We integrated Eq.~\eqref{eq:motion} with an explicit Euler step for the drift and Gaussian
Brownian increments. The primary calculation uses local diffusion units with
\(\kappa=\sigma^2\) and \(\Delta t=10^{-3}\); the supplemental numerical
check also uses \(\Delta t=5\times10^{-4}\). At each step it: (i) evaluates
the drift from the current geometry and active edge state; (ii) applies the
deterministic and Brownian increments and wraps positions on the torus; (iii)
samples each label's replacement indicator with probability
\(p=1-e^{-\Delta t/\Theta}\); (iv) clears all edges incident to replaced
labels; (v) samples replacement positions prescribed by each rule and inserts
new agents; and (vi) recomputes pair geometry and updates edge states according
to the activation and deletion thresholds. The bounded relocation sampler uses the displacement length
\(\ell_{\rm move}=R\sqrt{U_1}\in[0,R]\) and angle
\(\phi=2\pi U_2\in[0,2\pi)\), where \(U_1\) and \(U_2\) are independent
uniform random variables on \([0,1]\). Every replacement rule consumes the
same two random numbers per replacement event, including replacement at the
same position, so the subsequent random number streams remain aligned.

\subsection{Prospective design, integrity, and statistical computation}

The primary design, power analysis, configurations, software, and tests were
fixed in an archived project protocol before simulation. Independent checks
validated all 76,800 trajectories before analysis and reconstructed the raw
and control variate estimates, simultaneous bootstrap intervals, and
critical values.
The Supplemental Material details the separate designs, numerical checks,
and inference procedures; all protocols and runtime records are retained
with the code.

\section*{Data availability}

Recorded simulation time series, configurations, derived tables, bootstrap
draws, and integrity metadata for the primary and supplemental experiments
are available from the corresponding author on request.

\section*{Code availability}

The simulator, analytic replacement transforms, configuration tools, batch
simulation code, analysis code, tests, and deterministic figure scripts are
available from the corresponding author upon reasonable request.

\bibliography{references}

\end{document}


\title{Supplemental Material for: Replacement distance controls growth at long wavelengths in a coevolving spatial network}
\author{Ruiwu Niu}
\email[Corresponding author: ]{rniu@hksyu.edu}
\affiliation{Department of Data Science and Digital Innovation, Hong Kong Shue Yan University, Hong Kong SAR, China}
\author{Ming Yuan}
\email[Email: ]{yuanming@sjtu.edu.cn}
\affiliation{State Key Laboratory of Mechanical System and Vibration, Shanghai Jiao Tong University, Shanghai 200240, China}
\author{Ying-kit Wong}
\email[Email: ]{ykwong@hksyu.edu}
\affiliation{Department of Data Science and Digital Innovation, Hong Kong Shue Yan University, Hong Kong SAR, China}
\date{October 12, 2026}
\maketitle

\section{Scope of the mathematical results}

The derivations use the following assumptions and approximations:

\begin{enumerate}
\item The bound on agent drift in Eq. (3) is exact for every graph and
   configuration of the discrete agent model.
\item The restoring rate of a compact cluster is exact when a common Euclidean
   lift reproduces all minimum-image displacements, every pair is connected, and every pair has unit
   edge weight, with no edges to agents outside the cluster.
\item The replacement operator is exact for the ensemble density of one agent.
   Its finite step multiplier at zero attraction is also exact for the
   implemented Brownian and Bernoulli update.
\item The contact profile written with modified Bessel functions is an exact
   continuous time result for one independent Brownian pair at zero attraction
   in a homogeneous
   background, with exponential agent replacement and an immediate edge
   activation check.
\item The dispersion relation at positive attraction is an annealed mean field
   approximation. It replaces the dynamic edge state and random local strength
   by their averages, linearizes the resulting reciprocal strength
   denominator, and neglects pair correlations caused by attraction and the
   finite response time of the edge network.
\end{enumerate}

The simulation experiment tests the exact baseline and the
approximate attraction response separately before evaluating their sum.

\paragraph{Uncertainty conventions.}
The independent unit is a complete matched simulation block. Simultaneous
intervals protect the explicitly stated family at its stated confidence
level; separate families are not jointly protected across tables, cohorts,
or analyses. Pointwise intervals are identified where used. A deterministic
theory value has no sampling interval. Each table retains the family size
and estimator needed to interpret its entries; computational definitions
are given in Sec.~\ref{sec:estimation}.

\section{Symmetric force normalization}

For an active edge, Eq. (1) gives \(0\le q_{ij}\le1\) and
\(r_{ij}<H\sigma\). The attractive drift of agent \(i\) in Eq. (2) is

\begin{equation}
b_i=\chi\sum_{j\ne i}
\frac{A_{ij}q_{ij}}{s_0+s_i+s_j}\,\Delta_{ij}.
\end{equation}

The triangle inequality and \(s_j\ge0\) give

\begin{equation}
\begin{aligned}
|b_i|
&\le \chi H\sigma
\sum_{j\ne i}\frac{A_{ij}q_{ij}}{s_0+s_i+s_j}\\
&\le \chi H\sigma
\frac{\sum_{j\ne i}A_{ij}q_{ij}}{s_0+s_i}
=\chi H\sigma\frac{s_i}{s_0+s_i}
\leq\chi H\sigma .
\end{aligned}
\end{equation}

For \(\chi>0\), the last inequality is strict. Thus increasing the degree
cannot make the attractive speed diverge. Equation
(S2) gives the bound stated in Eq. (3). Because
the denominator is symmetric in the two endpoints and
\(\Delta_{ji}=-\Delta_{ij}\), every pair contributes equal and opposite lifted
drift increments.

Consider \(n_c\in\{2,\ldots,N\}\) agents that form a complete active subgraph,
have \(q_{ij}=1\) for every pair, have no edges to agents outside the group,
and admit a common Euclidean lift on the torus whose pair differences
coincide with every minimum-image displacement within the group. Every agent then has
\(s_i=n_c-1\). Define the lifted centroid by
\(\bar X=n_c^{-1}\sum_{j=1}^{n_c}X_j\). It follows that

\begin{equation}
\sum_{j\ne i}(X_j-X_i)=n_c(\bar X-X_i),
\end{equation}

and hence the attractive drift satisfies

\begin{equation}
b_i=-a_{n_c}(X_i-\bar X),\qquad
a_{n_c}=\frac{\chi n_c}{s_0+2(n_c-1)}.
\end{equation}

For \(s_0=1\), \(a_{n_c}\to\chi/2\) as \(n_c\to\infty\). Without normalization,
the corresponding rate is \(\chi n_c\). Equation (S4) motivates the
normalization: cluster stiffness cannot dominate the comparison among
replacement rules merely because cluster population increases.

\section{Exact replacement transport}

Let \(n(\boldsymbol x,t_{\rm phys})\equiv
\rho(\boldsymbol x,t_{\rm phys})\) be the ensemble density in Eq. (6), with
\(n:\mathbb T_L^2\times[0,\infty)\to[0,\infty)\) and
\(\int_{\mathbb T_L^2}n\,d^2\boldsymbol x=N\). Let \(K\) be the probability law
of a displacement \(\boldsymbol\xi\in\mathbb T_L^2\). During an infinitesimal
physical time increment \(dt_{\rm phys}>0\), each agent is replaced with
probability \(\mu\,dt_{\rm phys}+o(dt_{\rm phys})\). To leading order, the
expected density removed at \(\boldsymbol x\) is
\(\mu n(\boldsymbol x,t_{\rm phys})dt_{\rm phys}\), whereas replacements arrive
after displacements drawn from \(K\). Here \(o(dt_{\rm phys})\) denotes terms
that vanish faster than \(dt_{\rm phys}\). Dividing by \(dt_{\rm phys}\) and
taking the limit gives

\begin{equation}
\left.\partial_{t_{\rm phys}}n(\boldsymbol x,t_{\rm phys})\right|_{\rm ren}
=\mu\left[\int_{\mathbb T_L^2}
n(\boldsymbol x-\boldsymbol\xi,t_{\rm phys})K(d\boldsymbol\xi)
-n(\boldsymbol x,t_{\rm phys})\right]
=\mu(K*n-n).
\end{equation}

This operator preserves \(\int n\,d^2\boldsymbol x=N\), although it transports
density over space. For
\(\boldsymbol m\in\mathbb Z^2\), define
\(\boldsymbol Q_{\boldsymbol m}=2\pi\boldsymbol m/L\),
\(Q_{\boldsymbol m}=|\boldsymbol Q_{\boldsymbol m}|\), and
\(u=Q_{\boldsymbol m}\sigma\). For a selected mode, write
\(\boldsymbol Q=\boldsymbol Q_{\boldsymbol m}\). We use the Fourier coefficients
\(\widehat n_{\boldsymbol Q}=\int n(\boldsymbol x)
e^{-\mathrm i\boldsymbol Q\cdot\boldsymbol x}\,d^2\boldsymbol x\) and
\(\widehat K(\boldsymbol Q)=\int e^{-\mathrm i\boldsymbol Q\cdot
\boldsymbol\xi}K(d\boldsymbol\xi)\). Then

\begin{equation}
\left.\partial_{t_{\rm phys}}\widehat n_{\boldsymbol Q}\right|_{\rm ren}
=\mu[\widehat K(\boldsymbol Q)-1]\widehat n_{\boldsymbol Q}.
\end{equation}

After introducing the dimensionless time
\(t=\kappa t_{\rm phys}/\sigma^2\), the contribution to the dimensionless growth
rate is \(\Theta^{-1}[\widehat K(u)-1]\), as stated in Eq. (7). The three
transforms in Eq. (8), specialized here to the original radius
\(R=5\sigma\), are

\begin{equation}
\widehat K_{\rm same}(u)=1,
\qquad
\widehat K_{\rm bounded}(u)=\frac{2J_1(5u)}{5u},
\qquad
\widehat K_{\rm global}(u)=0\quad(\boldsymbol m\ne\boldsymbol0).
\end{equation}

The bounded expression is defined at \(u=0\) by its continuous value one. The
zero mode always has \(\widehat K(0)=1\), including for global
relocation. Thus the spatial mean is conserved. The apparent discontinuity of
the global branch is a property of the finite torus: every nonzero Fourier mode is
damped at rate \(\mu\), while the exactly uniform mode is unchanged.

Equations (S5) and (S6) use physical time. From this point through Eq. (S9),
\(\Delta t>0\) denotes a dimensionless time step. The simulator first applies a
Brownian step and then independently replaces an agent with probability
\(p=1-e^{-\Delta t/\Theta}\in(0,1)\). The mean multiplier
\(M_K(u)\in\mathbb R\) of a mode is

\begin{equation}
M_K(u)=e^{-u^2\Delta t}
\{1+p[\widehat K(u)-1]\}.
\end{equation}

For the parameter values used here,
\(1+p[\widehat K(u)-1]>0\), so this multiplier gives the exact implemented rate
at zero attraction,

\begin{equation}
\lambda_{0,K}^{\Delta t}(u)
=-u^2+\Delta t^{-1}
\log\{1+p[\widehat K(u)-1]\}.
\end{equation}

Equation (S9) gives the exact zero attraction baseline used in main text
Eq. (9).

\section{Contact survival for one Brownian pair}

In the stationary continuous time reference at zero attraction, the relative
coordinate of two independently diffusing agents has generator
\(2\kappa\nabla^2\). Replacement of either endpoint
deletes the edge, so an active pair is lost at rate \(2\mu\). In the annulus
\(\sigma<r<H\sigma\), the density of active pairs has no source. Dividing by the
uniform marginal of relative position and writing
\(\zeta=r/\sigma\in[0,\infty)\) gives the boundary value problem in Eq. (10),

\begin{equation}
h_\Theta''(\zeta)+\zeta^{-1}h_\Theta'(\zeta)-\Theta^{-1}h_\Theta(\zeta)=0,
\qquad h_\Theta(1)=1,\quad h_\Theta(H)=0.
\end{equation}

Here \(h_\Theta:[0,\infty)\to[0,1]\) is the conditional probability that the
edge is active, and primes denote derivatives with respect to \(\zeta\). The
factor of two in relative diffusion and the factor of two in pair loss cancel.
With \(\gamma=\Theta^{-1/2}>0\), and with
\(I_0:[0,\infty)\to\mathbb R\) and
\(K_0:(0,\infty)\to\mathbb R\) denoting the modified Bessel functions of
order zero of the first and second kind,

\begin{equation}
h_\Theta(\zeta)=
\frac{I_0(\gamma H)K_0(\gamma \zeta)-K_0(\gamma H)I_0(\gamma \zeta)}
{I_0(\gamma H)K_0(\gamma)-K_0(\gamma H)I_0(\gamma)},
\qquad 1<\zeta<H.
\end{equation}

We complete the conditional profile with \(h_\Theta(\zeta)=1\) for
\(0\le\zeta\le1\) and \(h_\Theta(\zeta)=0\) for \(\zeta\ge H\). The boundary \(h_\Theta(1)=1\) follows from immediate activation whenever
a replacement appears within the inner contact disk.

The displacement rule does not enter Eq. (S10) because every replacement event
clears the incident edge, and each rule that is invariant under translation
preserves a uniform density. This pair calculation does not assume that
correlations between edges and density are independent of the displacement rule
at positive attraction.

\section{Derivation of the annealed attraction response}

The exact renewal calculation in Sec. S3 used the total density
\(n=N\varrho\), where
\(\varrho:\mathbb T_L^2\times[0,\infty)\to[0,\infty)\) is the probability
density of one indexed slot and
\(\int_{\mathbb T_L^2}\varrho\,d^2\boldsymbol x=1\). This symbol is distinct from
the replacement probability \(p\) in Eq. (S8). The force on a tagged agent
depends on the other \(N-1\) agents. We therefore use the density that excludes
the tagged agent,
\begin{equation*}
n_-(\boldsymbol x,t_{\rm phys})=(N-1)\varrho(\boldsymbol x,t_{\rm phys})
=\frac{N-1}{N}n(\boldsymbol x,t_{\rm phys}),
\qquad \int_{\mathbb T_L^2}n_-\,d^2\boldsymbol x=N-1.
\end{equation*}
Multiplication by the constant \((N-1)/N\) leaves every modal growth rate
unchanged; it only ensures that the mean strength counts possible neighbors
rather than the tagged agent itself.

Define the radial mean force weight

\begin{equation}
\psi_\Theta(\zeta)=h_\Theta(\zeta)\widetilde q(\zeta),
\qquad \widetilde q(\zeta)=q(\sigma\zeta).
\end{equation}

Both \(\psi_\Theta\) and \(\widetilde q\) map \([0,\infty)\) to \([0,1]\).
The radial transform of \(\psi_\Theta\) in two dimensions is

\begin{equation}
F_\Theta(u)=2\pi\int_0^H
\zeta\psi_\Theta(\zeta)J_0(u\zeta)\,d\zeta.
\end{equation}

This is the transform in main text Eq. (11), with
\(F_\Theta:[0,\infty)\to\mathbb R\).

The annealed closure replaces the binary dynamic edge field by
\(\psi_\Theta\). Let \(\boldsymbol z\in\mathbb R^2\) denote a lifted relative
displacement. Because \(\psi_\Theta(|\boldsymbol z|/\sigma)=0\) for
\(|\boldsymbol z|\ge H\sigma\) and \(H\sigma<L/2\), its support has a unique
Euclidean lift; the density fields below are extended periodically. Define the
dimensionless mean strength field
\(s:\mathbb T_L^2\times[0,\infty)\to[0,\infty)\) and the physical drift field
\(v:\mathbb T_L^2\times[0,\infty)\to\mathbb R^2\) by

\begin{equation}
s(\boldsymbol x,t_{\rm phys})=\int_{\mathbb R^2}
\psi_\Theta(|\boldsymbol z|/\sigma)
n_-(\boldsymbol x+\boldsymbol z,t_{\rm phys})\,d^2\boldsymbol z,
\end{equation}

\begin{equation}
v(\boldsymbol x,t_{\rm phys})=\chi\int_{\mathbb R^2}
\frac{\psi_\Theta(|\boldsymbol z|/\sigma)\boldsymbol z\,
n_-(\boldsymbol x+\boldsymbol z,t_{\rm phys})}
{s_0+s(\boldsymbol x,t_{\rm phys})
+s(\boldsymbol x+\boldsymbol z,t_{\rm phys})}\,d^2\boldsymbol z,
\end{equation}

\begin{equation}
\partial_{t_{\rm phys}} n_-=\kappa\nabla^2n_- -\nabla\!\cdot(n_-v)
+\mu(K*n_- -n_-).
\end{equation}

Equation (S16) is written in physical time. For the homogeneous state
\(n_-=\rho_-=(N-1)/L^2>0\),

\begin{equation}
a=\rho_-\sigma^2=\eta/\pi>0,
\qquad \bar s=aF_\Theta(0)\ge0,
\qquad d_0=s_0+2\bar s>0.
\end{equation}

At the simulation values \(\eta=6\), \(H=2\), \(s_0=1\), and \(\Theta=5\),
the quadrature gives
\begin{equation}
\begin{aligned}
a&\simeq1.90986, & F_\Theta(0)&\simeq5.46669,\\
\bar s&\simeq10.4406, & d_0&\simeq21.8812.
\end{aligned}
\end{equation}

For a nonzero wavevector \(\boldsymbol Q\), represent a real perturbation by
\(n_-=\rho_-+\delta n_{\boldsymbol Q}
e^{\mathrm i\boldsymbol Q\cdot\boldsymbol x}+\mathrm{c.c.}\), where
\(\delta n_{\boldsymbol Q}\in\mathbb C\) is small and \(\mathrm{c.c.}\) denotes
the complex conjugate term. Analyzing one Fourier component at first order gives

\begin{equation}
\delta s_{\boldsymbol Q}=\sigma^2F_\Theta(u)\delta n_{\boldsymbol Q},
\end{equation}

and radial symmetry gives

\begin{equation}
\int_{\mathbb R^2}\psi_\Theta(|\boldsymbol z|/\sigma)\boldsymbol z
e^{\mathrm i\boldsymbol Q\cdot\boldsymbol z}\,d^2\boldsymbol z
=-\mathrm i\sigma^3F_\Theta'(u)\widehat{\boldsymbol Q},
\qquad \widehat{\boldsymbol Q}=\frac{\boldsymbol Q}{|\boldsymbol Q|}.
\end{equation}

The unit vector \(\widehat{\boldsymbol Q}\) is used only for
\(\boldsymbol Q\ne\boldsymbol0\); the vector integral vanishes at the zero mode.
Expanding the denominator of Eq. (S15) yields three terms at first order. The
direct density perturbation in the numerator supplies \(d_0^{-1}\). The
perturbation of \(s(\boldsymbol x,t_{\rm phys})\) multiplies an odd unmodulated
vector integral and therefore vanishes. The perturbation of
\(s(\boldsymbol x+\boldsymbol z,t_{\rm phys})\) carries the phase
\(e^{\mathrm i\boldsymbol Q\cdot\boldsymbol z}\) and supplies
\(-aF_\Theta(u)d_0^{-2}\). Substitution into
\(-\nabla\cdot(n_-v)\), followed by scaling time with
\(\sigma^2/\kappa\), gives

\begin{equation}
G_\Theta(u)=
a\,u[-F_\Theta'(u)]
\left[d_0^{-1}-aF_\Theta(u)d_0^{-2}\right].
\end{equation}

Thus \(G_\Theta:[0,\infty)\to\mathbb R\). Radial symmetry gives
\(F_\Theta'(0)=0\), so \(G_\Theta(0)=0\) and the zero mode remains conserved.

This result is Eq. (12) of the main text. Together with the finite step baseline
in Eq. (9), it gives the approximation in main text Eq. (13):

\begin{equation}
\Lambda_K^{\Delta t}(u)
\simeq \lambda_{0,K}^{\Delta t}(u)+\mathrm{Pe}\,G_\Theta(u).
\end{equation}

Equation (S22) contains no fitted parameter, but its second term is not exact
for the agent process. In particular, it assumes that the edge field responds
instantaneously to the density perturbation. The simulation data show that the
response scale is accurate to a few percent over the selected long and
intermediate modes, while the total rate can differ appreciably when the
negative baseline and positive response nearly cancel.

\section{Numerical implementation and verification}

The primary simulation parameters, defined in Eqs. (4) and (5), were

\begin{equation}
N=256,\quad L=1,\quad \eta=6,\quad H=2,\quad s_0=1,
\quad \Theta=5,\quad \mathrm{Pe}\in\{0,10.5\}.
\end{equation}

They imply
\(\sigma=\sqrt{6/(255\pi)}\simeq0.08654\),
\(\kappa=\sigma^2\), and a bounded relocation radius
\(R=5\sigma\simeq0.4327<L/2\). The primary step size was \(10^{-3}\), the final time was
0.6, and the record spacing was 0.01. The four tested modes were
\((1,0),(2,0),(2,1),(4,0)\), with scaled wavenumbers
\(0.543764,1.087528,1.215893,2.175056\).
Initial positions followed the density in Eq. (14), and the recorded projected
amplitude was the observable in Eq. (15).

The fixed update order was: compute attraction; apply drift and Brownian
increments; wrap positions; sample replacement indicators; clear all incident
edges; sample new positions and assign new agent labels; recompute geometry;
and update edge states according to the activation and deletion thresholds.
Each replacement rule consumed the same two
replacement uniforms per event, including replacement at the same position. This kept
the five random streams aligned across paired conditions.

Before the main simulations, verification tests checked the empirical Fourier
transform of each relocation sampler, the exact one step multiplier at zero
attraction, removal of edges incident to replaced agents, replacement at the
same position, contact synchronization within the same step, periodic wrapping
of bounded displacements, alignment of the five common random number streams,
and bitwise agreement of the global implementation with the reference
implementation. The simulation matrix
contained 3,200 independent seed blocks and 24 conditions per block, or 76,800
trajectories in total.

\section{Estimation and simultaneous inference}
\label{sec:estimation}

For each mode, the raw estimate fits the ensemble projected amplitude to the
model in main text Eq. (16),

\begin{equation}
\overline A(t)=c e^{\lambda t},\qquad c\ge0,
\end{equation}

by bounded nonlinear least squares with \(-50<\lambda<50\), using the recorded
time points satisfying \(t\le t_{\rm fit}(u)\). The dimensionless fit window is

\begin{equation}
t_{\rm fit}(u)=\min\left[0.6,
\frac{\log 2}{u^2+\Theta^{-1}}\right].
\end{equation}

The primary estimate at positive attraction uses the paired control variate
with coefficient fixed to one, as in main text Eq. (17):

\begin{equation}
Z_\ell(t)=A_\ell^P(t)-
\left[A_\ell^0(t)-A_\ell^0(0)e^{\lambda_{0,K}^{\Delta t}(u)t}\right],
\qquad \ell\in\{1,\ldots,3200\}.
\end{equation}

Under the exact mean identity at zero attraction, the bracket has zero
expectation, so
\(\mathbb E[Z_\ell(t)]=\mathbb E[A_\ell^P(t)]\). The control variate changes
precision, not the target mean curve, and it does not use the approximation at
positive attraction in Eq. (S22).

The response estimator specified before the simulations fits

\begin{equation}
\overline{A^P(t)-A^0(t)}=
a_0\left[e^{[\lambda_{0,K}^{\Delta t}(u)+\Delta\lambda_K]t}
-e^{\lambda_{0,K}^{\Delta t}(u)t}\right].
\end{equation}

Here \(a_0=\overline{A^0(0)}\ge0\) and
\(\Delta\lambda_K\in\mathbb R\) is the fitted growth increment caused by
attraction. It is compared with
\(\Delta\lambda_{\rm th}=\mathrm{Pe}G_\Theta(u)\), obtained from Eq. (12)
without fitting.

Let \(B_{\rm boot}=10{,}000\). Each bootstrap draw
\(b\in\{1,\ldots,B_{\rm boot}\}\) resampled the 3,200 complete seed blocks once
and reused the same multinomial count vector for every rule, mode, attraction
condition, time point, fit, and contrast. For a simultaneous family
\(\{\theta_j:j=1,\ldots,J\}\subset\mathbb R\), where
\(J\in\{1,2,\ldots\}\), let
\(\widehat\theta_j\) be the estimate from the original sample,
\(\theta_{bj}^*\) its estimate in draw \(b\), and \(s_j>0\) its bootstrap
standard deviation. Define

\begin{equation}
M_b=\max_{1\le j\le J}
\left|\frac{\theta_{bj}^*-\widehat\theta_j}{s_j}\right|.
\end{equation}

For confidence level \(C\in(0,1)\), let \(q_C\) be the empirical \(C\) quantile
of the jointly finite \(M_b\) values, evaluated with the sample quantile
convention \nolinkurl{method="higher"}. The simultaneous interval is
\([\widehat\theta_j-q_C s_j,\widehat\theta_j+q_C s_j]\). A family was
declared not met if any constituent statistic, or the family jointly, was
finite in fewer than 99\% of draws. Recorded time points were never treated as
independent statistical units.

\section{Complete short time results}

\subsection{Calibration at zero attraction}

All 12 simultaneous intervals contain the exact rate in Eq. (9). The common
95\% critical value for the maximum standardized deviation was 2.720433.

\begin{table}[htbp]
\centering
\caption{Exact rates and raw estimates at zero attraction.}
\fontsize{10}{12}\selectfont
\setlength{\tabcolsep}{4pt}
\renewcommand{\arraystretch}{1.22}
\begin{tabular}{lccccc}
\toprule
\textbf{Rule} & \textbf{Mode} & \textbf{Exact} & \textbf{Estimate} & \shortstack{\textbf{Simultaneous}\\\textbf{95\% low}} & \shortstack{\textbf{Simultaneous}\\\textbf{95\% high}} \\
\midrule
Same position & (1,0) & -0.29568 & -0.31516 & -0.37878 & -0.25154 \\
Same position & (2,0) & -1.18272 & -1.16092 & -1.32728 & -0.99456 \\
Same position & (2,1) & -1.47840 & -1.46248 & -1.67384 & -1.25112 \\
Same position & (4,0) & -4.73087 & -4.79889 & -5.43180 & -4.16597 \\
Bounded relocation & (1,0) & -0.43156 & -0.44900 & -0.52715 & -0.37085 \\
Bounded relocation & (2,0) & -1.40804 & -1.41750 & -1.61089 & -1.22412 \\
Bounded relocation & (2,1) & -1.69553 & -1.77790 & -2.01791 & -1.53789 \\
Bounded relocation & (4,0) & -4.93661 & -5.02924 & -5.68287 & -4.37561 \\
Global & (1,0) & -0.49568 & -0.50364 & -0.58904 & -0.41823 \\
Global & (2,0) & -1.38272 & -1.33932 & -1.52656 & -1.15207 \\
Global & (2,1) & -1.67840 & -1.72148 & -1.95559 & -1.48737 \\
Global & (4,0) & -4.93087 & -5.00910 & -5.66275 & -4.35544 \\
\bottomrule
\end{tabular}
\end{table}

\subsection{Primary growth signs and estimator comparison at positive attraction}

The control variate intervals form one simultaneous 95\% family with critical
value 2.601210. The raw diagnostic intervals form a separate simultaneous 95\%
family with critical value 2.728514 and do not enter the primary joint
criterion.

\begin{table}[htbp]
\centering
\caption{Theory, control variate estimates, and unadjusted estimates for the nine primary conditions.}
\fontsize{10}{12}\selectfont
\setlength{\tabcolsep}{3pt}
\renewcommand{\arraystretch}{1.35}
\begin{tabular}{>{\raggedright\arraybackslash}p{69pt}cccc>{\raggedright\arraybackslash}p{57pt}}
\toprule
\textbf{Rule} & \textbf{Mode} & \shortstack{\textbf{Theory}\\\textbf{total}} & \shortstack{\textbf{Control variate}\\\textbf{estimate [sim. 95\%]}} & \shortstack{\textbf{Raw estimate}\\\textbf{[sim. 95\%]}} & \textbf{Required sign resolved?} \\
\midrule
Same position & (1,0) & 0.08821 & \shortstack{0.06999\\{}[0.05802, 0.08196]} & \shortstack{0.05363\\{}[-0.00278, 0.11003]} & Yes \\
Bounded relocation & (1,0) & -0.04767 & \shortstack{-0.06058\\{}[-0.07408, -0.04708]} & \shortstack{-0.07558\\{}[-0.14482, -0.00634]} & Yes \\
Global & (1,0) & -0.11179 & \shortstack{-0.12387\\{}[-0.13772, -0.11002]} & \shortstack{-0.13055\\{}[-0.20677, -0.05433]} & Yes \\
Same position & (2,0) & 0.32730 & \shortstack{0.24347\\{}[0.19851, 0.28843]} & \shortstack{0.25523\\{}[0.14525, 0.36520]} & Yes \\
Bounded relocation & (2,0) & 0.10197 & \shortstack{0.03781\\{}[-0.01210, 0.08773]} & \shortstack{0.03532\\{}[-0.09287, 0.16350]} & No: unresolved \\
Global & (2,0) & 0.12730 & \shortstack{0.05657\\{}[0.00799, 0.10515]} & \shortstack{0.08520\\{}[-0.03892, 0.20932]} & Yes \\
Same position & (4,0) & -0.75092 & \shortstack{-0.41379\\{}[-0.53842, -0.28916]} & \shortstack{-0.45628\\{}[-0.88228, -0.03028]} & Yes \\
Bounded relocation & (4,0) & -0.95666 & \shortstack{-0.60959\\{}[-0.73682, -0.48237]} & \shortstack{-0.66089\\{}[-1.10258, -0.21920]} & Yes \\
Global & (4,0) & -0.95092 & \shortstack{-0.60840\\{}[-0.73583, -0.48098]} & \shortstack{-0.66368\\{}[-1.10620, -0.22115]} & Yes \\
\bottomrule
\end{tabular}
\end{table}

The theory and empirical point estimates have the same sign in all nine
conditions. The theoretical numerical value lies inside only two of the nine
control variate simultaneous intervals: bounded relocation and global relocation at
mode \((1,0)\).

\subsection{Contrasts at the longest wavelength}

For \(\boldsymbol m=(1,0)\), let
\(\lambda_{\rm same}\), \(\lambda_{\rm bounded}\), and
\(\lambda_{\rm global}\) denote the paired control variate estimates under the
three replacement rules. Their contrasts form one simultaneous 95\% family
with critical value 2.214875:

\begin{equation}
\lambda_{\rm same}-\lambda_{\rm bounded}
=0.13057\ [0.12421,0.13692],
\end{equation}

\begin{equation}
\lambda_{\rm same}-\lambda_{\rm global}
=0.19385\ [0.18745,0.20026].
\end{equation}

Because all three rules clear incident edges, these contrasts isolate the
effect of relocating replacement density.

\subsection{Attraction response and equivalence among replacement rules}

\begin{table}[htbp]
\centering
\caption{Paired attraction responses entering the quantitative comparison.}
\fontsize{10}{12}\selectfont
\setlength{\tabcolsep}{4pt}
\renewcommand{\arraystretch}{1.35}
\begin{tabular}{lccccc}
\toprule
\textbf{Rule} & \textbf{Mode} & \textbf{Theory} & \shortstack{\textbf{Empirical response}\\\textbf{[sim. 95\%]}} & \textbf{Empirical/theory} & \shortstack{\textbf{Absolute}\\\textbf{error (\%)}} \\
\midrule
Global & (1,0) & 0.38388 & \shortstack{0.37524\\{}[0.36391, 0.38658]} & 0.9775 & 2.25 \\
Same position & (1,0) & 0.38388 & \shortstack{0.37050\\{}[0.36059, 0.38040]} & 0.9651 & 3.49 \\
Bounded relocation & (1,0) & 0.38388 & \shortstack{0.37459\\{}[0.36355, 0.38563]} & 0.9758 & 2.42 \\
Global & (2,0) & 1.51002 & \shortstack{1.45451\\{}[1.41404, 1.49497]} & 0.9632 & 3.68 \\
Same position & (2,0) & 1.51002 & \shortstack{1.44397\\{}[1.40643, 1.48152]} & 0.9563 & 4.37 \\
Bounded relocation & (2,0) & 1.51002 & \shortstack{1.45974\\{}[1.41830, 1.50118]} & 0.9667 & 3.33 \\
\bottomrule
\end{tabular}
\end{table}

All six empirical responses are below the prediction evaluated without fitting.
Their mean absolute percentage error (MAPE) is 3.2566\%, with bootstrap 95\% upper bound
4.3755\%. Thus the response error is small but systematic in direction.

The six normalized contrasts among replacement rules form one simultaneous 90\%
family with critical value 2.329746. Their interval endpoints range from -0.025293 to
0.012994, wholly inside the prespecified equivalence band [-0.15,0.15]. This
supports statistical equivalence within the prespecified margin over the
tested conditions, not exact equality of the three processes with attraction.

\begin{table}[htbp]
\centering
\caption{Normalized differences between the paired attraction responses.}
\fontsize{10}{12}\selectfont
\setlength{\tabcolsep}{4pt}
\renewcommand{\arraystretch}{1.35}
\begin{tabular}{c>{\raggedright\arraybackslash}p{135pt}ccc}
\toprule
\textbf{Mode} & \textbf{Contrast} & \textbf{Estimate} & \shortstack{\textbf{Simultaneous}\\\textbf{90\% interval}} & \shortstack{\textbf{Inside}\\\textbf{[-0.15,0.15]?}} \\
\midrule
(1,0) & Same position minus bounded relocation & -0.010651 & [-0.023495, 0.002193] & Yes \\
(1,0) & Same position minus global & -0.012367 & [-0.025293, 0.000560] & Yes \\
(1,0) & Bounded relocation minus global & -0.001715 & [-0.015189, 0.011758] & Yes \\
(2,0) & Same position minus bounded relocation & -0.010440 & [-0.020939, 0.000059] & Yes \\
(2,0) & Same position minus global & -0.006975 & [-0.016362, 0.002411] & Yes \\
(2,0) & Bounded relocation minus global & 0.003465 & [-0.006064, 0.012994] & Yes \\
\bottomrule
\end{tabular}
\end{table}

\subsection{Nearby intermediate mode selected in advance}

Mode \((2,1)\) was selected before the simulations to test how the initial
response varies across intermediate wavelengths. It did not enter the primary joint
sign criterion. Its three control variate intervals are positive, whereas the
raw bounded and global intervals are unresolved.

\begin{table}[htbp]
\centering
\caption{Diagnostic for the nearby intermediate mode.}
\fontsize{10}{12}\selectfont
\setlength{\tabcolsep}{4pt}
\renewcommand{\arraystretch}{1.35}
\begin{tabular}{lcccc}
\toprule
\textbf{Rule} & \shortstack{\textbf{Theory}\\\textbf{total}} & \shortstack{\textbf{Control variate}\\\textbf{estimate [sim. 95\%]}} & \shortstack{\textbf{Raw estimate}\\\textbf{[sim. 95\%]}} & \shortstack{\textbf{Response}\\\textbf{error (\%)}} \\
\midrule
Same position & 0.37878 & \shortstack{0.31071\\{}[0.26481, 0.35662]} & \shortstack{0.31026\\{}[0.17206, 0.44846]} & 2.78 \\
Bounded relocation & 0.16165 & \shortstack{0.07626\\{}[0.02551, 0.12700]} & \shortstack{0.03127\\{}[-0.12718, 0.18973]} & 3.44 \\
Global & 0.17878 & \shortstack{0.11880\\{}[0.06922, 0.16837]} & \shortstack{0.08913\\{}[-0.06439, 0.24266]} & 2.41 \\
\bottomrule
\end{tabular}
\end{table}

None of the three theoretical total rates lies inside its corresponding
control variate interval. This result illustrates cancellation sensitivity:
an error of only a few percent in the attraction response can noticeably shift
a small residual total rate.

\subsection{Combined result}

\begin{table}[htbp]
\centering
\caption{Components of the prespecified joint criterion.}
\fontsize{10}{12}\selectfont
\setlength{\tabcolsep}{4pt}
\renewcommand{\arraystretch}{1.25}
\begin{tabular}{@{}>{\raggedright\arraybackslash}p{0.23\textwidth}>{\raggedright\arraybackslash}p{0.45\textwidth}>{\raggedright\arraybackslash}p{0.27\textwidth}@{}}
\toprule
\textbf{Component} & \textbf{Requirement} & \textbf{Result} \\
\midrule
Baseline at zero attraction & 12 simultaneous intervals cover Eq. (S9) & Satisfied, 12/12 \\
Primary growth signs & All nine control variate intervals lie on their required side of zero & Not met, 8/9; bounded \((2,0)\) unresolved \\
Contrasts at the longest wavelength & Both contrasts between same position and displaced replacement are positive & Satisfied, 2/2 \\
Response accuracy & \(\mathrm{MAPE}\leq10\%\) and bootstrap 95\% upper bound \(\leq15\%\) & Satisfied, 3.2566\% and 4.3755\% \\
Equivalence among rules & Six simultaneous 90\% intervals inside [-0.15,0.15] & Satisfied, 6/6 \\
Combined criterion & Every preceding component is met & Not met \\
\bottomrule
\end{tabular}
\end{table}

The bounded \((2,0)\) interval straddles zero. All nine point estimates
have their prescribed signs, and eight are resolved by the simultaneous
intervals.

\section{Retrospective checks of the initial response}
\label{sec:retrospective_response}

The following analyses use the original 3,200 independent seed blocks but were chosen after the original analysis. They test the sensitivity of the initial response and do not replace the criteria specified before the simulations.

\subsection{Fit window and raw amplitude curves}

The sign contrast at the longest wavelength persists when the fit window changes. We refit all 12 conditions at $\mathrm{Pe}=10.5$, comprising three replacement rules and four wave vectors $\bm m\in\{(1,0),(2,0),(2,1),(4,0)\}$. The dimensionless time interval is $[0,f t_{\rm fit}]$ with $f\in\{1/2,3/4,1\}$, or $[0.1,t_{\rm fit}]$, where $t_{\rm fit}$ is the original mode dependent upper limit. Only recorded points, spaced by $0.01$, enter each fit. The exact null correction in $Z_\ell(t)$ always uses the initial amplitude $A^0_\ell(0)$, including when $t=0$ is excluded from the fit.

Each fit uses the original free amplitude exponential estimator and the same 10,000 multinomial bootstrap draws with seed 2026081317. Each draw resamples 3,200 complete seed blocks, preserving pairing across rules, attraction strengths, and times. A maximum standardized bootstrap deviation gives simultaneous 95\% intervals across the 12 rates within each window. These intervals are not simultaneous across the four window choices.

\begin{table*}[t]
\caption{Fit window sensitivity at $\mathrm{Pe}=10.5$. Entries give the rate estimate and its simultaneous 95\% interval, rounded to three decimal places. The bounded rule uses $R=5\sigma$. Each column is a separate family of 12 comparisons. In the final column, the $(4,0)$ fits use only five time points from $0.10$ to $0.14$.}
\label{tab:window_sensitivity}
\footnotesize
\setlength{\tabcolsep}{3pt}
\begin{tabular}{llcccc}
\toprule
Mode & Rule & Half window & Three quarters & Original & $t\ge0.1$ \\
\midrule
$(1,0)$ & Same position & \shortstack{$0.078$\\$[0.067,0.088]$} & \shortstack{$0.075$\\$[0.064,0.087]$} & \shortstack{$0.070$\\$[0.058,0.082]$} & \shortstack{$0.067$\\$[0.052,0.081]$} \\
$(1,0)$ & Bounded & \shortstack{$-0.055$\\$[-0.067,-0.044]$} & \shortstack{$-0.057$\\$[-0.070,-0.044]$} & \shortstack{$-0.061$\\$[-0.075,-0.047]$} & \shortstack{$-0.063$\\$[-0.079,-0.046]$} \\
$(1,0)$ & Global & \shortstack{$-0.120$\\$[-0.132,-0.109]$} & \shortstack{$-0.121$\\$[-0.134,-0.108]$} & \shortstack{$-0.124$\\$[-0.138,-0.109]$} & \shortstack{$-0.126$\\$[-0.143,-0.109]$} \\
$(2,0)$ & Same position & \shortstack{$0.279$\\$[0.240,0.318]$} & \shortstack{$0.257$\\$[0.213,0.300]$} & \shortstack{$0.243$\\$[0.197,0.290]$} & \shortstack{$0.230$\\$[0.175,0.286]$} \\
$(2,0)$ & Bounded & \shortstack{$0.065$\\$[0.023,0.108]$} & \shortstack{$0.049$\\$[0.001,0.097]$} & \shortstack{$0.038$\\$[-0.014,0.090]$} & \shortstack{$0.027$\\$[-0.035,0.089]$} \\
$(2,0)$ & Global & \shortstack{$0.086$\\$[0.044,0.128]$} & \shortstack{$0.071$\\$[0.024,0.118]$} & \shortstack{$0.057$\\$[0.006,0.107]$} & \shortstack{$0.044$\\$[-0.016,0.105]$} \\
$(2,1)$ & Same position & \shortstack{$0.349$\\$[0.300,0.398]$} & \shortstack{$0.329$\\$[0.275,0.383]$} & \shortstack{$0.311$\\$[0.254,0.368]$} & \shortstack{$0.295$\\$[0.226,0.364]$} \\
$(2,1)$ & Bounded & \shortstack{$0.126$\\$[0.073,0.180]$} & \shortstack{$0.101$\\$[0.042,0.161]$} & \shortstack{$0.076$\\$[0.013,0.139]$} & \shortstack{$0.054$\\$[-0.023,0.130]$} \\
$(2,1)$ & Global & \shortstack{$0.154$\\$[0.102,0.207]$} & \shortstack{$0.138$\\$[0.080,0.196]$} & \shortstack{$0.119$\\$[0.057,0.180]$} & \shortstack{$0.103$\\$[0.028,0.177]$} \\
$(4,0)$ & Same position & \shortstack{$-0.470$\\$[-0.577,-0.363]$} & \shortstack{$-0.430$\\$[-0.547,-0.313]$} & \shortstack{$-0.414$\\$[-0.543,-0.284]$} & \shortstack{$-0.392$\\$[-0.638,-0.145]$} \\
$(4,0)$ & Bounded & \shortstack{$-0.667$\\$[-0.776,-0.558]$} & \shortstack{$-0.627$\\$[-0.746,-0.508]$} & \shortstack{$-0.610$\\$[-0.742,-0.478]$} & \shortstack{$-0.582$\\$[-0.836,-0.329]$} \\
$(4,0)$ & Global & \shortstack{$-0.659$\\$[-0.769,-0.550]$} & \shortstack{$-0.622$\\$[-0.741,-0.502]$} & \shortstack{$-0.608$\\$[-0.741,-0.476]$} & \shortstack{$-0.587$\\$[-0.842,-0.333]$} \\
\bottomrule
\end{tabular}
\end{table*}

All 12 point estimates retain their signs across the four windows (Table~\ref{tab:window_sensitivity}). More specifically, the longest mode has a positive interval for replacement at the same position and negative intervals for bounded and global replacement in every window. Sign resolution at intermediate wavelengths is less stable: all 12 intervals exclude zero in the half and three quarters windows, 11 do so in the original window, and nine do so after excluding $t<0.1$. Statistical resolution is therefore weaker after the earliest records are excluded.

The fitted rates also depend on the observation interval. For bounded replacement at $\bm m=(2,1)$, the half window estimate exceeds the original estimate by $0.050$, with a simultaneous 95\% interval $[0.003,0.097]$; discarding $t<0.1$ shifts it by $-0.023$, with interval $[-0.044,-0.001]$. These differences use paired bootstrap draws and separate families of 12 rate differences for each window comparison. The five point late window for $\bm m=(4,0)$ spans only $0.04$ time units and provides a limited diagnostic. Together, these results show a reproducible reversal of the broad pattern and a dependence of fitted rates on the observation interval.

\begin{figure*}[t]
\centering
\includegraphics[width=\textwidth]{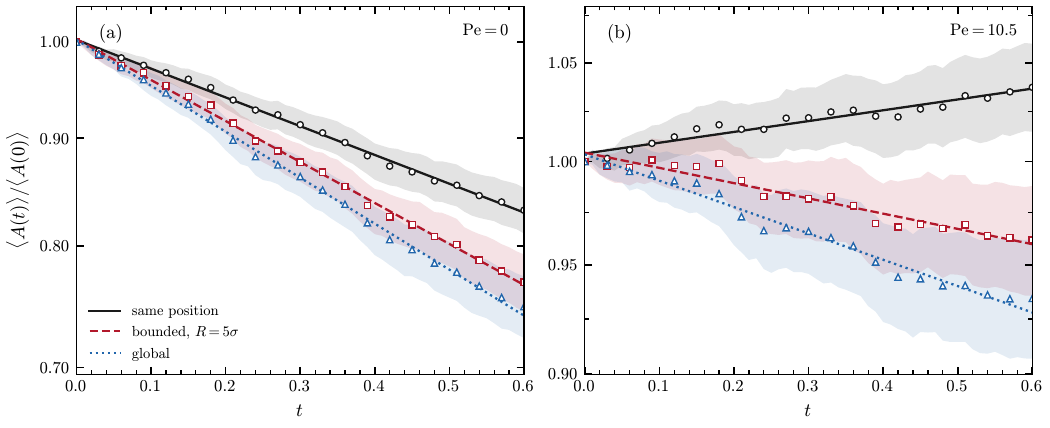}
\caption{Raw ensemble amplitude of the longest nonzero mode, $\bm m=(1,0)$, (a) without attraction and (b) at $\mathrm{Pe}=10.5$. The horizontal axis is dimensionless time $t$; the logarithmic vertical axis is $\overline A(t)/\overline A(0)$, the ensemble mean projected amplitude divided by its ensemble mean at the initial time. Symbols show the recorded means every $0.03$ time units for readability; every $0.01$ record enters the fit. Black circles denote replacement at the same position, red squares bounded replacement with $R=5\sigma$, and blue triangles global replacement. Lines are separate raw free amplitude exponential fits over $0\le t\le0.6$, divided by $\overline A(0)$; no control variate rate is imposed on a raw curve. Shading gives pointwise 95\% percentile intervals for the normalized raw mean from 10,000 bootstrap resamples of the 3,200 complete seed blocks. These are not simultaneous confidence bands.}
\label{fig:raw_initial_amplitude}
\end{figure*}

Figure~\ref{fig:raw_initial_amplitude} shows the underlying longest wavelength signal without a control variate. Without attraction, the mean amplitude decreases under all three rules. With attraction, replacement at the same position permits growth, whereas bounded and global replacement retain decay. These raw curves provide a direct view of the measured signal; the paired control variate supplies the more precise rate comparisons used in the main text.

\subsection{Local rates from overlapping fit windows}
\label{sec:local_fits}

To display the time dependence without reducing it to four broad windows,
we fit the same ensemble control variate to a free amplitude exponential
over windows of width $0.1$, advancing the left endpoint by $0.02$.
Each window contains 11 original records and lies entirely inside that
mode's original fit interval. The fitted rate
$\widehat\lambda_{\rm loc}(t_{\rm mid})\in\mathbb R$ is assigned to the
window midpoint $t_{\rm mid}\ge0.05$. It is an effective rate over a finite
interval, not an instantaneous derivative. The fit is nonlinear least
squares on the original amplitude scale, not a regression on logarithms.
The control variate retains the actual $t=0$ amplitude in every window.

The same 10,000 whole-block resamples of the 3,200 original blocks supply
pointwise percentile 95\% intervals. A point curve must remain positive,
and at least 99\% of bootstrap curves must be positive with finite fitted
rates; otherwise the window is retained as nonestimable. All 198 windows
meet these numerical criteria. Individual signed trajectories are neither
clipped nor replaced by their absolute values. Adjacent windows overlap,
so their estimates are correlated and the intervals are not a simultaneous
test of a trend. The shortest mode permits only three complete windows.

\begin{figure}[htbp]
\centering
\includegraphics[width=\textwidth]{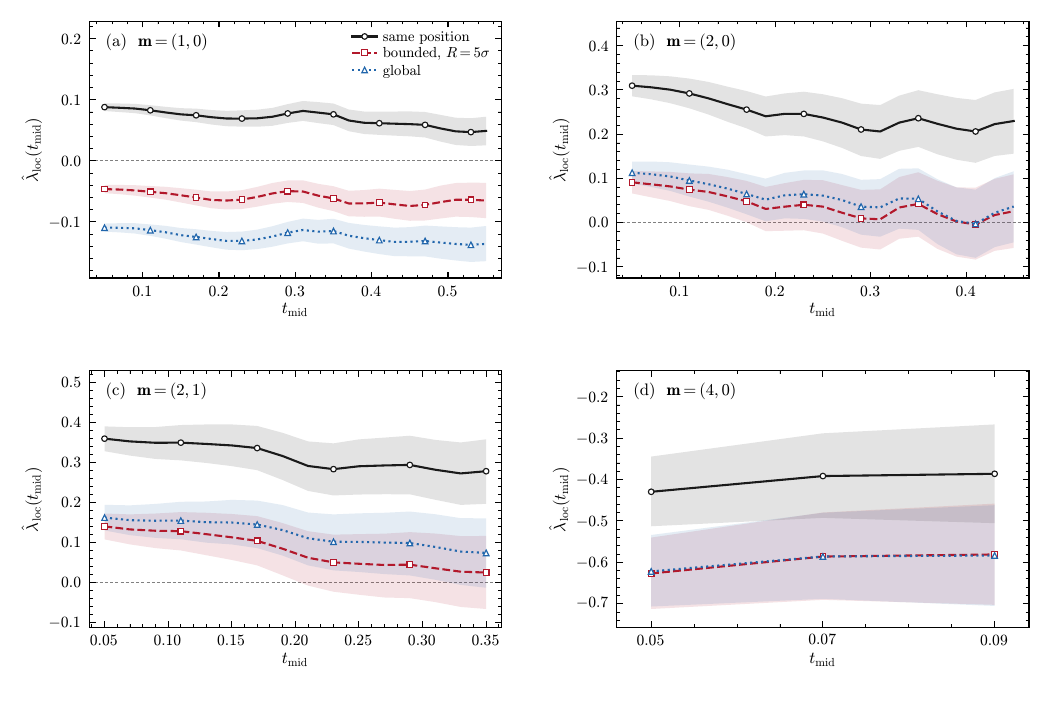}
\caption{Exploratory local exponential fits to the original ensemble
control variate curves. Panels (a)--(d) correspond to
$\boldsymbol m=(1,0),(2,0),(2,1),(4,0)$. The horizontal axis is the
midpoint of an overlapping window of width $0.1$; the vertical axis is
its fitted rate. Black circles, red squares, and blue triangles denote
replacement at the same position, within the $R=5\sigma$ disk, and globally.
Lines connect estimates as guides. Shading shows pointwise 95\% bootstrap
intervals, not simultaneous bands. Each panel uses only windows within
its original mode-specific fit interval; panel (d) therefore contains only
three estimates per rule. Changes along a curve describe dependence on
the fitting interval rather than an instantaneous stability eigenvalue.}
\label{fig:local_rates}
\end{figure}

\subsection{Bounded replacement can damp more strongly than global replacement}

A bounded displacement does not always preserve a density mode better than a global displacement. For a uniform disk of radius $R=5\sigma$, the displacement transform is $\widehat K_R(u)=2J_1(5u)/(5u)$, where $u=Q\sigma>0$ is the dimensionless wave number, $Q$ is the magnitude of the spatial wave vector, and $J_1$ is the Bessel function of the first kind and order one. Its first negative interval is approximately $0.766<u<1.403$. The modes $(2,0)$ and $(2,1)$ lie in this interval. A negative transform means that the replacement displacement reverses the phase contribution on average, so its damping can exceed that of the global rule, whose transform vanishes for every nonzero mode.

Table~\ref{tab:disk_global_contrasts} compares the fitted bounded and global rates in the original fit windows. The zero attraction estimates are new retrospective contrasts, with four simultaneous intervals formed from the same paired bootstrap draws. The positive attraction estimates retain the original simultaneous intervals over all 12 pairwise rule contrasts (three pairs at four modes); the four displayed entries are a subset of that family. The negative baseline prediction is therefore distinguished from its statistical resolution: the zero attraction intervals at the intermediate and short wavelengths all include zero, while the positive attraction contrast at $(2,1)$ is negative. The latter is consistent with the negative transform, but does not by itself identify a source of the discrepancy between the attraction closure and simulation.

\begin{table*}[t]
\caption{Bounded minus global fitted rate, with simultaneous 95\% intervals. Negative differences indicate a lower net growth rate under bounded replacement. The last column is the exact difference of the discrete zero attraction baseline rates. The two measured columns belong to different simultaneous families, as specified in the text. Values are rounded to three decimal places. The endpoint marked with an asterisk is positive before rounding, $3.1\times10^{-4}$.}
\label{tab:disk_global_contrasts}
\begin{tabular}{lccc}
\toprule
Mode & $\mathrm{Pe}=0$ & $\mathrm{Pe}=10.5$ & Exact baseline \\
\midrule
$(1,0)$ & $0.055\,[0.000^{*},0.109]$ & $0.063\,[0.055,0.072]$ & $0.064$ \\
$(2,0)$ & $-0.078\,[-0.160,0.004]$ & $-0.019\,[-0.042,0.004]$ & $-0.025$ \\
$(2,1)$ & $-0.056\,[-0.155,0.042]$ & $-0.043\,[-0.068,-0.017]$ & $-0.017$ \\
$(4,0)$ & $-0.020\,[-0.179,0.139]$ & $-0.001\,[-0.035,0.033]$ & $-0.006$ \\
\bottomrule
\end{tabular}
\end{table*}

\section{Independent supplemental experiments}
\label{sec:independent_extension}

These independent experiments test the replacement mechanism and numerical
sensitivity. Their separate archived protocol was fixed after the primary
analysis and before production with new seeds. They used Python 3.13.14,
NumPy 2.3.5, and SciPy 1.18.0 on arm64 macOS. Their estimates use new
trajectories and are reported separately from the original joint criterion.

\subsection{Radius scan and inference}

The radius scan uses $N=256$, $L=1$, $\eta=6$, $H=2$, $s_0=1$,
$\Theta=5$, $\epsilon=0.08$, $\Delta t=0.001$, and
$\bm m=(1,0)$. The disk radii are
$c_R=R/\sigma\in\{1,2,3,3.5,4,4.5,5,5.5\}$; replacement at the same
position and global relocation provide two controls. Each of 800 independent
blocks contains all ten rules at $\mathrm{Pe}=0$ and $10.5$, for 16,000
trajectories. All trajectories end at $t=0.6$ and are recorded every $0.01$.
The largest radius is $R/L\simeq0.476<1/2$.

Within each block, the unchanged simulator shares its five random streams
across rules and attraction conditions. The initial configurations and
initial edges are identical within the block. Blocks use separate seeds.
Fits use the original free amplitude exponential estimator and its
mode dependent window. The control variate uses the exact null rate with
the appropriate disk radius and time step, and each trajectory's own initial
amplitude. Thus changing the radius does not insert the attraction closure
into the estimator.

We use 10,000 bootstrap resamples of complete matched blocks, with analysis
seed 2026100209. Maximum standardized deviations give a simultaneous 95\%
family of ten rates with attraction and a separate family of nine rate
contrasts with replacement at the same position. Subtracting a deterministic
contrast between null rates shifts both interval endpoints by the same amount.
No interpolation of a zero crossing is treated as an experimentally resolved
critical radius. Point estimates must be finite and at least 99\% of bootstrap
draws must be jointly finite within each family; otherwise the corresponding
analysis is reported as nonestimable. No additional sampling is triggered by
an unresolved result.

The rates are listed in Table~\ref{tab:radius_scan_rates} and shown in
main text Fig.~6. Replacement at the same position and disk radii
$c_R=1,2$ have positive simultaneous intervals. Radii $4,4.5,5,5.5$
and global relocation have negative intervals, while $3$ and $3.5$ are
unresolved. Every one of the nine contrasts with replacement at the same
position has a negative interval. After subtracting its exact baseline
contrast, each of the nine residual intervals contains zero; the residual
point estimates span $[-0.0022,0.0031]$. These residuals are consistent with zero at the reported precision.
They compare fitted control variate rates; main text Fig.~3(a) uses the
separate paired increment estimator.

\begin{table}[htbp]
\centering
\caption{Independent radius scan at $\bm m=(1,0)$ and $\mathrm{Pe}=10.5$, with 800 matched blocks. The control variate column has simultaneous 95\% intervals over all ten rules; the raw column has pointwise 95\% percentile intervals. The columns therefore differ both in estimator and in interval family. Values are rounded to three decimal places, except very small nonzero values, which retain their signs in scientific notation.}
\label{tab:radius_scan_rates}
\small
\setlength{\tabcolsep}{4pt}
\renewcommand{\arraystretch}{1.13}
\begin{tabular}{lcc}
\toprule
Replacement rule & Control variate [sim. 95\%] & Raw [pointwise 95\%] \\
\midrule
Same position & $0.072\,[0.051,0.094]$ & $0.087\,[0.009,0.164]$ \\
Disk, $R/\sigma=1$ & $0.063\,[0.041,0.085]$ & $0.070\,[-0.010,0.147]$ \\
Disk, $R/\sigma=2$ & $0.044\,[0.022,0.066]$ & $0.040\,[-0.043,0.120]$ \\
Disk, $R/\sigma=3$ & $0.016\,[-0.007,0.038]$ & $3.7\times10^{-4}\,[-0.088,0.086]$ \\
Disk, $R/\sigma=3.5$ & $-0.004\,[-0.028,0.019]$ & $-0.025\,[-0.117,0.064]$ \\
Disk, $R/\sigma=4$ & $-0.025\,[-0.049,-0.001]$ & $-0.050\,[-0.146,0.042]$ \\
Disk, $R/\sigma=4.5$ & $-0.046\,[-0.070,-0.022]$ & $-0.076\,[-0.175,0.019]$ \\
Disk, $R/\sigma=5$ & $-0.066\,[-0.091,-0.041]$ & $-0.099\,[-0.201,-0.001]$ \\
Disk, $R/\sigma=5.5$ & $-0.082\,[-0.108,-0.057]$ & $-0.118\,[-0.221,-0.018]$ \\
Global & $-0.126\,[-0.152,-0.100]$ & $-0.125\,[-0.232,-0.022]$ \\
\bottomrule
\end{tabular}
\end{table}

\subsection{Local pair structure during the fit window}

We record local structure for replacement at the same position, within the
$R=5\sigma$ disk, and globally, at both attraction strengths in all 800
blocks of the radius scan. For a distance cutoff $d_c\in(0,L/2)$, define
the disk averaged pair density ratio
\begin{equation*}
\overline g_{d_c}(t)=
\frac{L^2}{\pi d_c^2 \binom{N}{2}}
\left\langle\sum_{i<j}{\bf1}\{r_{ij}(t)<d_c\}\right\rangle,
\end{equation*}
where $\langle\cdot\rangle$ denotes the ensemble average over blocks and
$\mathbf1$ is the indicator function. This ratio is dimensionless and
nonnegative. The denominator is the expected pair
count for independent uniform positions. Unity denotes the independent uniform reference; the weak initial
modulation shifts the expected ratio at order $\epsilon^2$. This ratio
integrates the pair density over a disk. We record $d_c=\sigma$ and $d_c=0.3\sigma$, together
with the mean active degree
$\langle k(t)\rangle=N^{-1}\langle\sum_{i,j}A_{ij}(t)\rangle\in[0,N-1]$.

Curve bands are pointwise mean $\pm1.96$ standard errors across independent
blocks. Endpoint intervals are pointwise bootstrap percentile intervals;
contrasts between attraction conditions preserve the pairing. The independent replicate remains the complete block. These observables
measure reorganization of the prepared local background.

\begin{figure}[htbp]
\centering
\includegraphics[width=\textwidth]{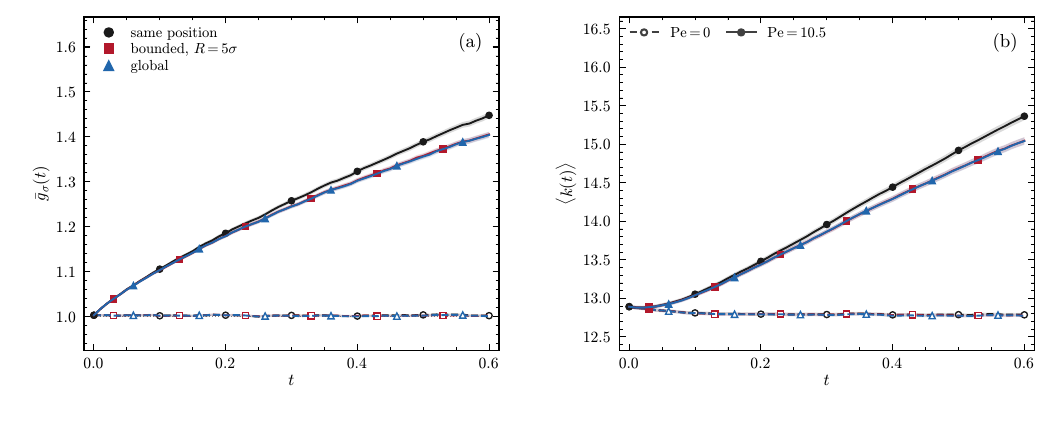}
\caption{Local pair structure during the initial response of the longest
mode. (a) Pair count within one contact radius, divided by its independent
uniform expectation, $\overline g_\sigma(t)$. (b) Mean active degree.
The horizontal axis is dimensionless time. Black circles denote replacement
at the same position, red squares the $R=5\sigma$ disk rule, and blue
triangles global relocation. Lines include every recorded time; symbols
are shown at distinct recorded times for the three rules to avoid overlap.
Solid curves with filled symbols use
$\mathrm{Pe}=10.5$; dashed curves with open symbols use $\mathrm{Pe}=0$.
Bands show the mean plus or minus $1.96$ standard errors over 800 independent
blocks and are descriptive pointwise intervals, not simultaneous bands.
The dotted horizontal line in (a) is the independent uniform reference;
the initial weak modulation changes it only at order $\epsilon^2$.}
\label{fig:local_structure}
\end{figure}

Figure~\ref{fig:local_structure} follows the local pair density and mean
active degree through the same observation interval. Without attraction,
$\overline g_\sigma(0.6)$ is between $1.001$ and $1.002$ across rules.
With attraction it reaches $1.448$, $1.405$, and $1.404$ for replacement
at the same position, within the disk, and globally. The corresponding
mean active degrees are $15.36$, $15.05$, and $15.04$, compared with about
$12.78$ without attraction. The smaller distance cutoff also shows an
excess of close pairs (Table~\ref{tab:structure_endpoints}). All these
changes occur while the longest density mode decays under the displaced
rules. They separate local contact accumulation from amplification of the
broad imposed density pattern.

\begin{table}[htbp]
\centering
\caption{Local structure at $t=0.6$ in the independent radius scan. Each estimate averages 800 independent blocks. The last column gives the paired difference between attraction strengths with a pointwise bootstrap 95\% interval. The two $\overline g$ observables are integrated pair-density ratios, not pointwise radial distribution functions. The initial modulation is $\epsilon=0.08$.}
\label{tab:structure_endpoints}
\small
\setlength{\tabcolsep}{4pt}
\renewcommand{\arraystretch}{1.13}
\begin{tabular}{llccc}
\toprule
Replacement & Observable & $\mathrm{Pe}=0$ & $\mathrm{Pe}=10.5$ & Difference [95\%] \\
\midrule
Same position & $\overline g_\sigma$ & $1.002$ & $1.448$ & $0.446\,[0.440,0.452]$ \\
Same position & $\overline g_{0.3\sigma}$ & $0.998$ & $1.783$ & $0.785\,[0.769,0.801]$ \\
Same position & $\langle k\rangle$ & $12.786$ & $15.363$ & $2.577\,[2.537,2.619]$ \\
Disk, $R=5\sigma$ & $\overline g_\sigma$ & $1.002$ & $1.405$ & $0.403\,[0.398,0.409]$ \\
Disk, $R=5\sigma$ & $\overline g_{0.3\sigma}$ & $0.996$ & $1.722$ & $0.726\,[0.711,0.742]$ \\
Disk, $R=5\sigma$ & $\langle k\rangle$ & $12.777$ & $15.046$ & $2.269\,[2.231,2.307]$ \\
Global & $\overline g_\sigma$ & $1.001$ & $1.404$ & $0.403\,[0.398,0.409]$ \\
Global & $\overline g_{0.3\sigma}$ & $0.997$ & $1.724$ & $0.727\,[0.711,0.743]$ \\
Global & $\langle k\rangle$ & $12.773$ & $15.041$ & $2.267\,[2.229,2.306]$ \\
\bottomrule
\end{tabular}
\end{table}

\subsection{Time step and perturbation checks}

A separate set of 400 independent blocks contains four targets: all three
rules at $\bm m=(1,0)$ and the $R=5\sigma$ disk rule at $(2,0)$.
Each target is simulated at both attraction strengths with the original
settings, half the time step, or half the initial perturbation. The resulting
24 trajectories per block give 9,600 trajectories. Each trajectory has the
same final time and recording grid.

The two grids are coupled without changing either marginal update. A coarse
Brownian increment is the sum of two independent fine increments. Its
replacement indicator is the logical OR of the two fine indicators, giving
exactly the coarse probability $1-\exp(-\Delta t/\Theta)$. If replacement
occurs, the uniforms associated with the last fine event give the coarse
replacement coordinates. Two fine replacement events therefore remain two
events on the fine grid but one on the coarse grid. This is a coupling of
the numerical schemes, not an assertion that their paths coincide.
The condition with half the perturbation amplitude uses its own initial density from rejection sampling
and initial mode amplitude. Its two attraction conditions are still paired.

The eight differences, comprising two numerical changes for each of four
targets, form one simultaneous 95\% bootstrap family. The fit window and
sampling grid are unchanged. We report the paired changes and their simultaneous intervals as a
sensitivity check. The sample size was fixed using resource and precision
considerations.

\begin{figure}[htbp]
\centering
\includegraphics[width=3.375in]{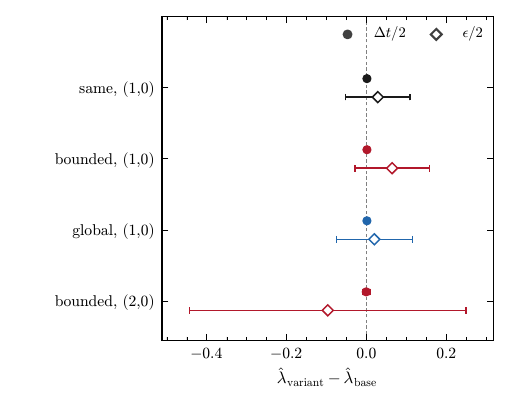}
\caption{Numerical sensitivity of the fitted control variate rates at
$\mathrm{Pe}=10.5$. Each row identifies a replacement rule and Fourier mode;
bounded relocation uses $R=5\sigma$. The horizontal axis is the rate with
one setting halved minus its rate at the original settings
$\Delta t=0.001$ and $\epsilon=0.08$. Filled circles halve the time step;
open diamonds halve the modulation amplitude. Horizontal bars are
simultaneous 95\% intervals over all eight paired contrasts, from 400
independent matched blocks. The vertical dashed line denotes no change.
The narrow time step intervals lie close to zero, whereas the amplitude
intervals remain too wide to demonstrate an amplitude-independent response.
The three longest-mode time step intervals are narrower than their symbols
at this scale.}
\label{fig:numerical_sensitivity}
\end{figure}

Figure~\ref{fig:numerical_sensitivity} separates the two checks. Halving
the time step changes the longest-mode rates by approximately
\(0.0003\) under each rule. Their joint intervals extend no farther than
\(-0.0025\) and \(0.0032\). For the disk's \((2,0)\) mode, the difference is
\(-0.0014\), with interval \([-0.0112,0.0084]\). The four tested targets show small sensitivity to this reduction of the
time step.

The amplitude comparisons have much lower precision. For the longest mode,
the difference is \(0.028\) with interval \([-0.053,0.108]\) at the same
position, \(0.063\) with interval \([-0.030,0.156]\) for the disk, and
\(0.019\) with interval \([-0.076,0.114]\) for global relocation.
The disk's \((2,0)\) difference is \(-0.098\), with interval
\([-0.443,0.248]\). Table~\ref{tab:robustness_rates}
shows that the half-amplitude disk condition does not resolve even the
longest-mode sign. Thus the available data do not establish independence
from perturbation amplitude or validate the limit of an infinitesimal
perturbation. All control variate bootstrap fits are finite. Two of the
10,000 secondary raw-null fits for the half-amplitude disk \((2,0)\)
condition are invalid; its percentile interval uses the 9,998 finite fits,
above the prespecified 99\% threshold. No block was replaced or added.

\begin{table}[htbp]
\centering
\caption{Rates in the independent numerical sensitivity study, with 400 matched blocks and $\mathrm{Pe}=10.5$. The original settings are $\Delta t=0.001$ and $\epsilon=0.08$. Both disk targets use $R=5\sigma$. Each entry is a control variate rate and a pointwise bootstrap 95\% interval; these are not the joint intervals for differences displayed in Fig.~\ref{fig:numerical_sensitivity}.}
\label{tab:robustness_rates}
\small
\setlength{\tabcolsep}{4pt}
\renewcommand{\arraystretch}{1.13}
\begin{tabular}{lccc}
\toprule
Rule and mode & Original settings & $\Delta t/2$ & $\epsilon/2$ \\
\midrule
Same position, $(1,0)$  & \shortstack{$0.061$\\$[0.036,0.086]$} & \shortstack{$0.061$\\$[0.036,0.087]$} & \shortstack{$0.089$\\$[0.036,0.142]$} \\
Disk, $(1,0)$  & \shortstack{$-0.076$\\$[-0.107,-0.046]$} & \shortstack{$-0.076$\\$[-0.106,-0.045]$} & \shortstack{$-0.013$\\$[-0.071,0.048]$} \\
Global, $(1,0)$  & \shortstack{$-0.133$\\$[-0.162,-0.104]$} & \shortstack{$-0.132$\\$[-0.162,-0.104]$} & \shortstack{$-0.114$\\$[-0.179,-0.050]$} \\
Disk, $(2,0)$  & \shortstack{$0.017$\\$[-0.096,0.125]$} & \shortstack{$0.016$\\$[-0.099,0.125]$} & \shortstack{$-0.081$\\$[-0.334,0.140]$} \\
\bottomrule
\end{tabular}
\end{table}

\subsection{Additional modes and paired amplitude comparison}
\label{sec:new_modes_amplitude}

The additional mode design fills gaps between the original sampled spatial
scales. It uses $\bm m\in\{(1,1),(2,2),(3,0),(3,1),(3,2)\}$,
all three replacement rules, and $\mathrm{Pe}\in\{0,10.5\}$.
The remaining parameters are $N=256$, $L=1$, $\eta=6$, $H=2$,
$s_0=1$, $\Theta=5$, $\epsilon=0.08$, and $\Delta t=0.001$.
Each trajectory ends at $t=0.6$, with records spaced by $0.01$.
Each of 3,200 new independent blocks contains all thirty conditions, giving
96,000 trajectories. These blocks share no seeds with the primary experiment,
the earlier supplemental experiments, or the amplitude comparison below.
The unchanged simulator shares its random streams within each block.

The original mode dependent fit window and both original rate estimators
are retained. The attraction response is estimated from the paired amplitude
difference with the measured initial amplitude held fixed, exactly as in the
primary analysis. It is not obtained by subtracting the null rate from a
control variate rate fit. Three separate simultaneous 95\% families contain
the fifteen rates with attraction, the fifteen raw rates without attraction,
and the fifteen attraction response ratios, respectively. Each family uses
10,000 resamples of complete blocks. The original primary and secondary
families remain separate from these new comparisons. New resolved signs can
bracket a sign change between sampled modes, but cannot change the original
nine-sign decision or identify a continuous critical wavenumber.

Tables~\ref{tab:x1_rates}--\ref{tab:x1_responses} report the complete
additional comparisons. Both displaced rules decay at $u=0.769$ and grow
at $u=1.538,1.631,1.720$ within the new rate family. The three intervals
at $u=1.961$ all include zero. The lower sampled sign bracket is therefore
$0.769$ to $1.216$ for bounded relocation, using the original secondary
positive mode, and $0.769$ to $1.088$ for global relocation, using an
original primary positive mode. The last resolved positive mode at $1.720$
and the original negative mode at $2.175$ supply the upper bracket under
all three rules. These discrete brackets combine signs reported in separate confidence
families. A joint confidence interval for a continuous band boundary would
require a separate analysis. All fifteen raw zero attraction intervals contain their exact
finite step rates. The new response ratios at $u=1.961$ exceed one by about
5\%, extending the scale dependent discrepancy seen in the original data.

A second design uses 1,600 other independent blocks. It compares
$\epsilon\in\{0.08,0.16\}$ at $\mathrm{Pe}\in\{0,10.5\}$ for
all three rules at $(1,0)$ and for the disk rule at $(2,0)$ and $(2,2)$.
Each block contains twenty conditions, for 32,000 trajectories. The two
amplitudes share random streams within a block; all other settings and fit
windows are unchanged. The five control variate rate differences,
$\widehat\lambda(0.16)-\widehat\lambda(0.08)$, form one simultaneous
95\% family based on 10,000 complete-block resamples. These comparisons
are separate from the older 400-block half-amplitude test. An interval
containing zero is not evidence of equivalence unless an equivalence
margin has also been specified.

There is a useful, but conditional, interpretation of the amplitude
comparison. Translating the initial cosine pattern by half a wavelength
changes the sign of $\epsilon$ and of its projected amplitude. The dynamics
are translation invariant and do not otherwise depend on $\epsilon$.
Consequently, the ensemble amplitude is odd in $\epsilon$. If the fit to
the normalized ensemble curve has a unique, smooth, nondegenerate interior
optimum in the zero-amplitude limit, its rate admits
an even expansion
$\lambda(\epsilon)=\lambda_{\rm lin}+c_2\epsilon^2+c_4\epsilon^4+\cdots$,
where $\lambda_{\rm lin},c_2,c_4\in\mathbb R$ refer to this preparation and
fit window, not to a stationary interacting state. Dividing
$\lambda(2\epsilon)-\lambda(\epsilon)$ by three estimates
$c_2\epsilon^2$ only when higher orders are negligible: its next term is
$5c_4\epsilon^4$, whereas the corresponding term in
$\lambda(\epsilon)-\lambda_{\rm lin}$ is $c_4\epsilon^4$.
We therefore interpret an interval divided by three as uncertainty in a
leading-order estimate, not as a rigorous bound on the zero-amplitude bias.
A resolved difference establishes amplitude dependence across the tested
range; it does not by itself determine the accuracy of the smaller amplitude.

\begin{figure}[htbp]
\centering
\includegraphics[width=3.375in]{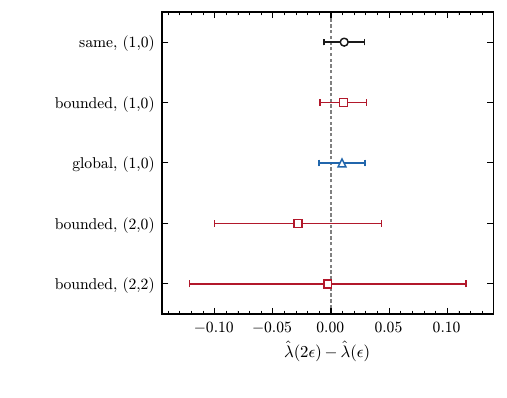}
\caption{Change in the fitted control variate rate on doubling the initial
modulation from $\epsilon=0.08$ to $0.16$ at $\mathrm{Pe}=10.5$.
The horizontal axis is the paired difference, and the vertical labels
identify the replacement rule and mode. Open symbols show estimates;
horizontal bars are simultaneous 95\% intervals across the five differences
from 1,600 independent blocks and 10,000 paired block resamples.
The vertical line is zero. All intervals contain zero, with substantially
less precision at the intermediate modes. These are fresh blocks, not the
400-block half-amplitude study in Fig.~\ref{fig:numerical_sensitivity}.}
\label{fig:double_amplitude}
\end{figure}

All five measured amplitude differences have intervals containing zero
(Table~\ref{tab:x2_contrasts}, Fig.~\ref{fig:double_amplitude}).
The longest-mode point differences range from $0.0095$ to $0.0114$;
their intervals lie within $[-0.0103,0.0307]$. For bounded $(2,0)$,
the difference is $-0.0284$, with interval $[-0.1000,0.0432]$.
For bounded $(2,2)$, it is $-0.0027$, with interval
$[-0.1215,0.1160]$. Conditional leading
corrections and their uncertainty appear in the last table column.
Their largest absolute interval endpoints are about $0.01$ for the
longest mode and $0.03$--$0.04$ for the two intermediate modes.
The latter precision is insufficient to establish an amplitude independent
rate near a weak sign change.

\begin{table}[htbp]
\centering
\caption{Additional modes at $\mathrm{Pe}=10.5$, from 3,200 new independent matched blocks. The control variate estimates and intervals form one simultaneous 95\% family of fifteen rates. The theory column is Eq.~(13), evaluated without fitted parameters. These conditions do not enter the original primary criterion.}
\label{tab:x1_rates}
\small
\setlength{\tabcolsep}{5pt}
\renewcommand{\arraystretch}{1.12}
\begin{tabular}{cclcc}
\toprule
$\boldsymbol m$ & $u$ & Rule & Theory & Estimate [sim. 95\%] \\
\midrule
$(1,1)$ & 0.769 & Same position & 0.177 & \shortstack{$0.132$\\$[0.106,0.157]$} \\
$(1,1)$ & 0.769 & Bounded & -0.024 & \shortstack{$-0.050$\\$[-0.079,-0.021]$} \\
$(1,1)$ & 0.769 & Global & -0.023 & \shortstack{$-0.054$\\$[-0.082,-0.026]$} \\
$(2,2)$ & 1.538 & Same position & 0.396 & \shortstack{$0.358$\\$[0.277,0.439]$} \\
$(2,2)$ & 1.538 & Bounded & 0.206 & \shortstack{$0.182$\\$[0.097,0.267]$} \\
$(2,2)$ & 1.538 & Global & 0.196 & \shortstack{$0.168$\\$[0.082,0.253]$} \\
$(3,0)$ & 1.631 & Same position & 0.350 & \shortstack{$0.371$\\$[0.284,0.459]$} \\
$(3,0)$ & 1.631 & Bounded & 0.162 & \shortstack{$0.215$\\$[0.124,0.306]$} \\
$(3,0)$ & 1.631 & Global & 0.150 & \shortstack{$0.187$\\$[0.096,0.279]$} \\
$(3,1)$ & 1.720 & Same position & 0.276 & \shortstack{$0.295$\\$[0.197,0.393]$} \\
$(3,1)$ & 1.720 & Bounded & 0.088 & \shortstack{$0.126$\\$[0.024,0.228]$} \\
$(3,1)$ & 1.720 & Global & 0.076 & \shortstack{$0.112$\\$[0.010,0.214]$} \\
$(3,2)$ & 1.961 & Same position & -0.117 & \shortstack{$0.113$\\$[-0.007,0.233]$} \\
$(3,2)$ & 1.961 & Bounded & -0.313 & \shortstack{$-0.093$\\$[-0.216,0.030]$} \\
$(3,2)$ & 1.961 & Global & -0.317 & \shortstack{$-0.078$\\$[-0.202,0.046]$} \\
\bottomrule
\end{tabular}
\end{table}

\begin{table}[htbp]
\centering
\caption{Calibration without attraction for the additional modes, using the same 3,200 blocks as Table~\ref{tab:x1_rates}. The raw rates and intervals form a separate simultaneous 95\% family of fifteen estimates. The exact column uses the finite-step replacement and Brownian multiplier. All fifteen intervals contain their exact values.}
\label{tab:x1_baselines}
\small
\setlength{\tabcolsep}{5pt}
\renewcommand{\arraystretch}{1.12}
\begin{tabular}{cclcc}
\toprule
$\boldsymbol m$ & $u$ & Rule & Exact rate & Raw rate [sim. 95\%] \\
\midrule
$(1,1)$ & 0.769 & Same position & -0.591 & \shortstack{$-0.620$\\$[-0.719,-0.521]$} \\
$(1,1)$ & 0.769 & Bounded & -0.792 & \shortstack{$-0.794$\\$[-0.913,-0.674]$} \\
$(1,1)$ & 0.769 & Global & -0.791 & \shortstack{$-0.802$\\$[-0.921,-0.682]$} \\
$(2,2)$ & 1.538 & Same position & -2.365 & \shortstack{$-2.423$\\$[-2.761,-2.084]$} \\
$(2,2)$ & 1.538 & Bounded & -2.556 & \shortstack{$-2.642$\\$[-3.003,-2.280]$} \\
$(2,2)$ & 1.538 & Global & -2.565 & \shortstack{$-2.635$\\$[-2.993,-2.277]$} \\
$(3,0)$ & 1.631 & Same position & -2.661 & \shortstack{$-2.729$\\$[-3.097,-2.360]$} \\
$(3,0)$ & 1.631 & Bounded & -2.849 & \shortstack{$-2.869$\\$[-3.254,-2.484]$} \\
$(3,0)$ & 1.631 & Global & -2.861 & \shortstack{$-2.929$\\$[-3.318,-2.541]$} \\
$(3,1)$ & 1.720 & Same position & -2.957 & \shortstack{$-3.027$\\$[-3.450,-2.603]$} \\
$(3,1)$ & 1.720 & Bounded & -3.144 & \shortstack{$-3.190$\\$[-3.631,-2.749]$} \\
$(3,1)$ & 1.720 & Global & -3.157 & \shortstack{$-3.251$\\$[-3.701,-2.801]$} \\
$(3,2)$ & 1.961 & Same position & -3.844 & \shortstack{$-3.545$\\$[-4.079,-3.011]$} \\
$(3,2)$ & 1.961 & Bounded & -4.040 & \shortstack{$-3.735$\\$[-4.289,-3.181]$} \\
$(3,2)$ & 1.961 & Global & -4.044 & \shortstack{$-3.767$\\$[-4.320,-3.213]$} \\
\bottomrule
\end{tabular}
\end{table}

\begin{table}[htbp]
\centering
\caption{Paired attraction responses for the additional modes. Each response fits the paired amplitude difference with its measured initial amplitude fixed. It is not the control variate rate minus the exact baseline. Ratios divide by the positive theoretical attraction increment $\mathrm{Pe}\,G_\Theta(u)$ and have their own simultaneous 95\% family across fifteen conditions.}
\label{tab:x1_responses}
\small
\setlength{\tabcolsep}{5pt}
\renewcommand{\arraystretch}{1.12}
\begin{tabular}{cclccc}
\toprule
$\boldsymbol m$ & $u$ & Rule & Theory response & Paired response & Ratio [sim. 95\%] \\
\midrule
$(1,1)$ & 0.769 & Same position & 0.768 & 0.734 & \shortstack{$0.956$\\$[0.930,0.983]$} \\
$(1,1)$ & 0.769 & Bounded & 0.768 & 0.748 & \shortstack{$0.974$\\$[0.945,1.004]$} \\
$(1,1)$ & 0.769 & Global & 0.768 & 0.745 & \shortstack{$0.970$\\$[0.941,0.999]$} \\
$(2,2)$ & 1.538 & Same position & 2.762 & 2.744 & \shortstack{$0.994$\\$[0.969,1.018]$} \\
$(2,2)$ & 1.538 & Bounded & 2.762 & 2.757 & \shortstack{$0.998$\\$[0.973,1.024]$} \\
$(2,2)$ & 1.538 & Global & 2.762 & 2.753 & \shortstack{$0.997$\\$[0.971,1.022]$} \\
$(3,0)$ & 1.631 & Same position & 3.011 & 3.034 & \shortstack{$1.007$\\$[0.983,1.031]$} \\
$(3,0)$ & 1.631 & Bounded & 3.011 & 3.059 & \shortstack{$1.016$\\$[0.991,1.041]$} \\
$(3,0)$ & 1.631 & Global & 3.011 & 3.050 & \shortstack{$1.013$\\$[0.988,1.038]$} \\
$(3,1)$ & 1.720 & Same position & 3.232 & 3.254 & \shortstack{$1.007$\\$[0.982,1.032]$} \\
$(3,1)$ & 1.720 & Bounded & 3.232 & 3.270 & \shortstack{$1.012$\\$[0.986,1.038]$} \\
$(3,1)$ & 1.720 & Global & 3.232 & 3.267 & \shortstack{$1.011$\\$[0.985,1.037]$} \\
$(3,2)$ & 1.961 & Same position & 3.727 & 3.921 & \shortstack{$1.052$\\$[1.025,1.079]$} \\
$(3,2)$ & 1.961 & Bounded & 3.727 & 3.916 & \shortstack{$1.051$\\$[1.023,1.078]$} \\
$(3,2)$ & 1.961 & Global & 3.727 & 3.930 & \shortstack{$1.054$\\$[1.027,1.082]$} \\
\bottomrule
\end{tabular}
\end{table}

\begin{table}[htbp]
\centering
\caption{Paired amplitude comparison in 1,600 further independent blocks. Define $\delta_\epsilon=\hat\lambda(0.16)-\hat\lambda(0.08)$; the five differences form one simultaneous 95\% family. The last column is the same interval divided by three. It estimates the leading nonlinear correction at $\epsilon=0.08$ only if the even expansion of the fitted rate applies and fourth and higher orders are negligible; it is not a rigorous bound on zero-amplitude bias.}
\label{tab:x2_contrasts}
\small
\setlength{\tabcolsep}{5pt}
\renewcommand{\arraystretch}{1.12}
\begin{tabular}{clcc}
\toprule
$\boldsymbol m$ & Rule & $\delta_\epsilon$ [sim. 95\%] & $\delta_\epsilon/3$ [conditional] \\
\midrule
$(1,0)$ & Same position & \shortstack{$0.011$\\$[-0.006,0.029]$} & \shortstack{$0.0038$\\$[-0.0021,0.0096]$} \\
$(1,0)$ & Bounded & \shortstack{$0.011$\\$[-0.009,0.031]$} & \shortstack{$0.0035$\\$[-0.0032,0.0102]$} \\
$(1,0)$ & Global & \shortstack{$0.009$\\$[-0.010,0.029]$} & \shortstack{$0.0032$\\$[-0.0034,0.0098]$} \\
$(2,0)$ & Bounded & \shortstack{$-0.028$\\$[-0.100,0.043]$} & \shortstack{$-0.0095$\\$[-0.0333,0.0144]$} \\
$(2,2)$ & Bounded & \shortstack{$-0.003$\\$[-0.122,0.116]$} & \shortstack{$-0.0009$\\$[-0.0405,0.0387]$} \\
\bottomrule
\end{tabular}
\end{table}

\clearpage
\subsection{Frozen retrospective comparisons of early rates and replacement rules}
\label{sec:early_reanalysis}

The original local-window diagnostic motivated a common comparison of the
earliest window, $0\le t\le0.1$, across three existing cohorts. Before
performing this reanalysis, we fixed the eleven time points, estimators,
10,000 complete-block resamples, confidence families, and expectations.
The underlying trajectories had already been examined in earlier analyses.
This retrospective specification fixed the new analysis before its
computation, using only the existing trajectories.

For each condition, the early rate $\widehat\lambda_{\rm early}\in\mathbb R$
is the free amplitude exponential fit to the ensemble control variate
curve. The full rate $\widehat\lambda_{\rm full}\in\mathbb R$ uses the
original $t_{\rm fit}(u)$ and its recorded points. Both retain the original
initial amplitude in the exact null correction. The paired change is
$d=\widehat\lambda_{\rm full}-\widehat\lambda_{\rm early}\in\mathbb R$.
Each cohort has two separate simultaneous 95\% families, one for its early
rates and one for its changes: twelve entries each in the original
3,200-block cohort, fifteen in the additional 3,200-block cohort, and ten
in the 800-block radius cohort. The same bootstrap count vector is used
for all fits within a cohort, preserving pairing between windows.
All point fits and all bootstrap fits were finite, with positive early
mean curves. Tables~\ref{tab:early_original}--\ref{tab:early_radius} report
every condition. Rates over the full windows are reported in the earlier
cohort tables; their values are unchanged.

The frozen expectations for the additional modes were compatibility with
the theory at the twelve conditions with $u\le1.720$, early rates above
theory at the three conditions with $u=1.961$, and negative point changes
at the former twelve conditions. The first two expectations hold in
$12/12$ and $3/3$ conditions, respectively. The direction expectation
holds in $11/12$, but only $3/12$ changes have simultaneous intervals
below zero. No direction was expected for the shortest added mode; its
three changes have positive point estimates and unresolved signs. The
original cohort has $9/12$ compatible early intervals and $9/12$ negative
point changes, of which five are resolved. The radius cohort has $10/10$
compatible early intervals and $10/10$ negative point changes, none
resolved. These counts are reported separately, not combined into a pass
criterion. A theoretical value inside an interval is compatible with the estimate
at the available precision. For scale, dividing the early-minus-theory
point discrepancy by $\mathrm{Pe}\,G_\Theta(u)>0$ gives ranges of
$[-1.18,0.66]\%$ and $[-0.17,1.89]\%$ for the original and additional
long and intermediate conditions, $[8.07,8.29]\%$ and $[4.90,5.21]\%$
for their shortest modes, and $[0.68,1.93]\%$ for the radius cohort.

\begin{table}[htbp]
\caption{Original cohort: early rates and paired changes from 3,200 blocks.
The two interval columns are separate simultaneous 95\% families of twelve
entries. The theory is main text Eq. (13). Same denotes replacement at the
departure position; disk uses $R=5\sigma$. Values are rounded to three
decimals, with small nonzero endpoints retained in scientific notation.}
\label{tab:early_original}
\centering\small
\setlength{\tabcolsep}{4pt}
\begin{tabular}{llrcc}
\toprule
Mode & Rule & Theory & Early rate [sim. 95\%] & $d$ [sim. 95\%]\\
\midrule
$(1,0)$ & Same & $0.088$ & $0.088\,[0.080,0.096]$ & $-0.018\,[-0.031,-0.005]$ \\
$(1,0)$ & Disk & $-0.048$ & $-0.046\,[-0.055,-0.037]$ & $-0.014\,[-0.029,3.7\times10^{-4}]$ \\
$(1,0)$ & Global & $-0.112$ & $-0.109\,[-0.118,-0.100]$ & $-0.015\,[-0.030,2.7\times10^{-4}]$ \\
$(2,0)$ & Same & $0.327$ & $0.310\,[0.278,0.341]$ & $-0.066\,[-0.115,-0.018]$ \\
$(2,0)$ & Disk & $0.102$ & $0.091\,[0.058,0.124]$ & $-0.053\,[-0.106,-2.1\times10^{-4}]$ \\
$(2,0)$ & Global & $0.127$ & $0.113\,[0.080,0.146]$ & $-0.056\,[-0.108,-0.004]$ \\
$(2,1)$ & Same & $0.379$ & $0.359\,[0.318,0.400]$ & $-0.048\,[-0.105,0.008]$ \\
$(2,1)$ & Disk & $0.162$ & $0.140\,[0.097,0.183]$ & $-0.064\,[-0.125,-0.002]$ \\
$(2,1)$ & Global & $0.179$ & $0.162\,[0.119,0.205]$ & $-0.043\,[-0.103,0.017]$ \\
$(4,0)$ & Same & $-0.751$ & $-0.430\,[-0.544,-0.316]$ & $0.016\,[-0.041,0.073]$ \\
$(4,0)$ & Disk & $-0.957$ & $-0.627\,[-0.744,-0.510]$ & $0.017\,[-0.041,0.075]$ \\
$(4,0)$ & Global & $-0.951$ & $-0.622\,[-0.739,-0.505]$ & $0.013\,[-0.045,0.072]$ \\
\bottomrule
\end{tabular}
\end{table}

\begin{table}[htbp]
\caption{Additional mode cohort: early rates and paired changes from 3,200
blocks. The two interval columns are separate simultaneous 95\% families
of fifteen entries, not pooled with Table~\ref{tab:early_original}.
Notation and rounding follow that table.}
\label{tab:early_x1}
\centering\small
\setlength{\tabcolsep}{4pt}
\begin{tabular}{llrcc}
\toprule
Mode & Rule & Theory & Early rate [sim. 95\%] & $d$ [sim. 95\%]\\
\midrule
$(1,1)$ & Same & $0.177$ & $0.175\,[0.159,0.192]$ & $-0.044\,[-0.071,-0.017]$ \\
$(1,1)$ & Disk & $-0.024$ & $-0.020\,[-0.037,-0.002]$ & $-0.031\,[-0.061,-4.5\times10^{-4}]$ \\
$(1,1)$ & Global & $-0.023$ & $-0.020\,[-0.038,-0.003]$ & $-0.034\,[-0.063,-0.004]$ \\
$(2,2)$ & Same & $0.396$ & $0.405\,[0.340,0.470]$ & $-0.048\,[-0.118,0.023]$ \\
$(2,2)$ & Disk & $0.206$ & $0.226\,[0.158,0.293]$ & $-0.044\,[-0.117,0.030]$ \\
$(2,2)$ & Global & $0.196$ & $0.213\,[0.146,0.281]$ & $-0.046\,[-0.119,0.028]$ \\
$(3,0)$ & Same & $0.350$ & $0.378\,[0.307,0.448]$ & $-0.006\,[-0.078,0.065]$ \\
$(3,0)$ & Disk & $0.162$ & $0.210\,[0.138,0.283]$ & $0.004\,[-0.070,0.078]$ \\
$(3,0)$ & Global & $0.150$ & $0.196\,[0.124,0.268]$ & $-0.008\,[-0.083,0.066]$ \\
$(3,1)$ & Same & $0.276$ & $0.322\,[0.244,0.401]$ & $-0.027\,[-0.099,0.044]$ \\
$(3,1)$ & Disk & $0.088$ & $0.149\,[0.068,0.230]$ & $-0.023\,[-0.098,0.051]$ \\
$(3,1)$ & Global & $0.076$ & $0.131\,[0.050,0.213]$ & $-0.019\,[-0.093,0.055]$ \\
$(3,2)$ & Same & $-0.117$ & $0.067\,[-0.038,0.173]$ & $0.046\,[-0.027,0.118]$ \\
$(3,2)$ & Disk & $-0.313$ & $-0.130\,[-0.238,-0.022]$ & $0.037\,[-0.037,0.112]$ \\
$(3,2)$ & Global & $-0.317$ & $-0.122\,[-0.231,-0.014]$ & $0.045\,[-0.030,0.119]$ \\
\bottomrule
\end{tabular}
\end{table}

\begin{table}[htbp]
\caption{Radius cohort at $(1,0)$: early rates and paired changes from 800
blocks. The two interval columns are separate simultaneous 95\% families
of ten entries. Here $c_R=R/\sigma$ applies only to disk relocation.}
\label{tab:early_radius}
\centering\small
\setlength{\tabcolsep}{4pt}
\begin{tabular}{lrcc}
\toprule
Rule & Theory & Early rate [sim. 95\%] & $d$ [sim. 95\%]\\
\midrule
Same & $0.088$ & $0.091\,[0.078,0.104]$ & $-0.019\,[-0.041,0.003]$ \\
Disk, $c_R=1$ & $0.081$ & $0.084\,[0.070,0.097]$ & $-0.020\,[-0.043,0.002]$ \\
Disk, $c_R=2$ & $0.060$ & $0.064\,[0.051,0.078]$ & $-0.020\,[-0.043,0.002]$ \\
Disk, $c_R=3$ & $0.029$ & $0.033\,[0.019,0.047]$ & $-0.017\,[-0.040,0.006]$ \\
Disk, $c_R=3.5$ & $0.010$ & $0.015\,[0.001,0.029]$ & $-0.019\,[-0.043,0.005]$ \\
Disk, $c_R=4$ & $-0.009$ & $-0.004\,[-0.018,0.010]$ & $-0.021\,[-0.046,0.003]$ \\
Disk, $c_R=4.5$ & $-0.028$ & $-0.023\,[-0.037,-0.009]$ & $-0.023\,[-0.048,0.002]$ \\
Disk, $c_R=5$ & $-0.048$ & $-0.041\,[-0.055,-0.027]$ & $-0.025\,[-0.050,0.001]$ \\
Disk, $c_R=5.5$ & $-0.066$ & $-0.059\,[-0.073,-0.044]$ & $-0.024\,[-0.050,0.002]$ \\
Global & $-0.112$ & $-0.105\,[-0.119,-0.090]$ & $-0.021\,[-0.047,0.005]$ \\
\bottomrule
\end{tabular}
\end{table}

The second reanalysis compares replacement rules using the original
fixed initial amplitude estimator of the paired attraction increment
$\Delta\lambda_K$, not a fitted control variate rate minus its null.
For each pair of rules $K,K'$, the normalized difference is
$E_{K,K'}=(\Delta\lambda_K-\Delta\lambda_{K'})/
[\mathrm{Pe}\,G_\Theta(u)]\in\mathbb R$.
The fifteen differences across the five additional modes form one
simultaneous 90\% family. Their intervals all lie within $[-0.15,0.15]$
(Table~\ref{tab:p2_contrasts}). This margin comes from the original
protocol; an equivalence comparison was not preregistered for these
additional modes. The original two primary modes retain their six-contrast
90\% family, and the remaining two original modes are covered by the
existing twelve-contrast 95\% family over all four original modes. Thus
the nine-mode statement combines separate, explicitly scoped comparisons,
not a single simultaneous equivalence test across all modes.

\begin{table}[htbp]
\caption{Normalized paired attraction-response differences at the five
additional modes. All fifteen entries form one simultaneous 90\% family;
all intervals lie inside the inherited $[-0.15,0.15]$ margin. Differences
are left rule minus right rule; S, D, and G denote same position, disk,
and global replacement. The small nonzero endpoint retains its sign.}
\label{tab:p2_contrasts}
\centering\small
\begin{tabular}{clc}
\toprule
Mode & Pair & $E_{K,K'}$ [sim. 90\%]\\
\midrule
$(1,1)$ & S versus D & $-0.018\,[-0.033,-0.003]$\\
$(1,1)$ & S versus G & $-0.014\,[-0.027,-4.1\times10^{-4}]$\\
$(1,1)$ & D versus G & $0.004\,[-0.009,0.018]$\\
$(2,2)$ & S versus D & $-0.005\,[-0.012,0.003]$\\
$(2,2)$ & S versus G & $-0.003\,[-0.010,0.004]$\\
$(2,2)$ & D versus G & $0.002\,[-0.006,0.009]$\\
$(3,0)$ & S versus D & $-0.008\,[-0.015,-0.001]$\\
$(3,0)$ & S versus G & $-0.005\,[-0.012,0.002]$\\
$(3,0)$ & D versus G & $0.003\,[-0.004,0.010]$\\
$(3,1)$ & S versus D & $-0.005\,[-0.012,0.002]$\\
$(3,1)$ & S versus G & $-0.004\,[-0.011,0.003]$\\
$(3,1)$ & D versus G & $0.001\,[-0.006,0.007]$\\
$(3,2)$ & S versus D & $0.002\,[-0.005,0.008]$\\
$(3,2)$ & S versus G & $-0.002\,[-0.009,0.004]$\\
$(3,2)$ & D versus G & $-0.004\,[-0.010,0.003]$\\
\bottomrule
\end{tabular}
\end{table}

For the descriptive ordering, ``lower'' means a strictly smaller point
estimate, evaluated separately against each displaced rule. Same position
replacement is lower than disk at seven of nine modes, lower than global
at all nine, and lower than both at seven (Table~\ref{tab:response_ordering}).
The table describes point-estimate ordering; statistical comparisons are
reported in their respective interval families above.

\begin{table}[htbp]
\caption{Descriptive ordering across all nine modes. Columns report
normalized point differences from the original paired response estimator;
no new confidence family is attached to this ordering count.}
\label{tab:response_ordering}
\centering\small
\begin{tabular}{cccc}
\toprule
Mode & Same minus disk & Same minus global & Same below both\\
\midrule
$(1,0)$ & $-0.011$ & $-0.012$ & Yes\\
$(2,0)$ & $-0.010$ & $-0.007$ & Yes\\
$(2,1)$ & $0.007$ & $-0.004$ & No\\
$(4,0)$ & $-0.002$ & $-0.002$ & Yes\\
$(1,1)$ & $-0.018$ & $-0.014$ & Yes\\
$(2,2)$ & $-0.005$ & $-0.003$ & Yes\\
$(3,0)$ & $-0.008$ & $-0.005$ & Yes\\
$(3,1)$ & $-0.005$ & $-0.004$ & Yes\\
$(3,2)$ & $0.002$ & $-0.002$ & No\\
\bottomrule
\end{tabular}
\end{table}

\clearpage
\subsection{Exploratory closure using the evolving radial pair structure}
\label{sec:evolving_pairs}

Can measured pair structure improve the static closure? We use the first
200 blocks of the additional mode study, designated before production,
for both prediction and observation. At $t=0,0.1,\ldots,0.6$, exact
weighted pair sums define the real transform
\[
F(u,t)=\frac{L^2}{\mathcal P\sigma^2}
\left\langle\sum_{i<j}A_{ij}q_{ij}J_0(u\zeta_{ij})\right\rangle_{200},
\qquad \mathcal P=\frac{N(N-1)}{2},\quad \zeta_{ij}=\frac{r_{ij}}{\sigma}\ge0.
\]
The brackets average blocks; $u\ge0$ and $t\in[0,0.6]$. Differentiation
with respect to $u$ holds the pair ensemble fixed, and
$\bar s(t)=aF(0,t)\ge0$ is mean weighted strength. Substituting these
mean moments into main text Eq. (12), with $d_0(t)=s_0+2\bar s(t)>0$,
gives the approximate reconstructed response $G_{\rm rec}(u,t)\in\mathbb R$.
Thus averaging precedes the nonlinear substitution. The initial pair
reference includes the imposed modulation,
$h_\Theta(\zeta)[1+\epsilon^2J_0(u\zeta)/2]$, not just $h_\Theta$.

Linear interpolation of $G_{\rm rec}$ gives the positive predicted
amplitude $B(t)=\exp[\lambda_{0,K}^{\Delta t}(u)t+
\mathrm{Pe}\int_0^tG_{\rm rec}(u,v)\,dv]$, where $v$ is integration time.
We fit $B$ and the measured control variate curve over the same original
window and time grid. Table~\ref{tab:evolving_pair_rates} reports signed
rates, with prediction and measurement uncertainty obtained from the same
10,000 resamples of the 200 complete blocks.

The reconstruction gives no consistent improvement. Its descriptive mean
absolute rate error over the fifteen conditions is $0.180$, compared with
$0.068$ for the original closure. This substitution still omits directional
response, edge relaxation and correlations of endpoint strengths. All sixty comparisons
over the four original diagnostic windows, the paired attraction responses,
pair profiles, and bootstrap draws remain archived.

\begin{table}[htbp]
\caption{Signed growth rates for the exploratory pair reconstruction.
The original column is the unfitted main text Eq. (13); the other columns
give the reconstructed prediction and measured control variate rate over
the original full window. Both use the same 200 blocks. Brackets are
pointwise 95\% percentile intervals, including uncertainty of the measured
moments in the reconstruction, not simultaneous intervals. Same denotes
replacement at the departure position.}
\label{tab:evolving_pair_rates}
\centering
\small
\setlength{\tabcolsep}{4pt}
\begin{tabular}{llrcc}
\toprule
Mode & Rule & Original & Reconstructed [95\%] & Measured [95\%]\\
\midrule
$(1,1)$ & Same & $0.177$ & $0.096\,[0.094,0.098]$ & $0.128\,[0.052,0.204]$ \\
$(1,1)$ & Disk & $-0.024$ & $-0.102\,[-0.104,-0.100]$ & $-0.011\,[-0.094,0.073]$ \\
$(1,1)$ & Global & $-0.023$ & $-0.100\,[-0.102,-0.098]$ & $-0.052\,[-0.136,0.032]$ \\
$(2,2)$ & Same & $0.396$ & $0.225\,[0.218,0.232]$ & $0.336\,[0.082,0.583]$ \\
$(2,2)$ & Disk & $0.206$ & $0.039\,[0.032,0.045]$ & $0.143\,[-0.130,0.401]$ \\
$(2,2)$ & Global & $0.196$ & $0.029\,[0.022,0.035]$ & $0.184\,[-0.087,0.443]$ \\
$(3,0)$ & Same & $0.350$ & $0.179\,[0.173,0.185]$ & $0.446\,[0.185,0.698]$ \\
$(3,0)$ & Disk & $0.162$ & $-0.006\,[-0.012,0.001]$ & $0.305\,[0.034,0.567]$ \\
$(3,0)$ & Global & $0.150$ & $-0.018\,[-0.025,-0.012]$ & $0.278\,[0.004,0.539]$ \\
$(3,1)$ & Same & $0.276$ & $0.117\,[0.110,0.125]$ & $0.315\,[0.041,0.579]$ \\
$(3,1)$ & Disk & $0.088$ & $-0.067\,[-0.074,-0.059]$ & $0.180\,[-0.106,0.453]$ \\
$(3,1)$ & Global & $0.076$ & $-0.079\,[-0.086,-0.072]$ & $0.185\,[-0.096,0.460]$ \\
$(3,2)$ & Same & $-0.117$ & $-0.248\,[-0.256,-0.240]$ & $-0.026\,[-0.395,0.320]$ \\
$(3,2)$ & Disk & $-0.313$ & $-0.441\,[-0.449,-0.433]$ & $-0.246\,[-0.619,0.102]$ \\
$(3,2)$ & Global & $-0.317$ & $-0.446\,[-0.454,-0.438]$ & $-0.281\,[-0.666,0.073]$ \\
\bottomrule
\end{tabular}
\end{table}

\clearpage
\section{Earlier independent comparison with global relocation}

Before the experiment with three replacement rules, an earlier study conducted in two stages
compared global relocation at \(\Theta=5\) with a condition that had no agent
replacement at \(N=256\) and
\(\mathrm{Pe}=11\). The lowest tested mode changed from decay with replacement
to growth without replacement:

\begin{equation}
\lambda_{\rm global}=-0.09613,
\qquad
\lambda_{\rm no\ replacement}=0.11550,
\end{equation}

\begin{equation}
\lambda_{\rm no\ replacement}-\lambda_{\rm global}=0.21163,
\qquad 95\%\ \text{family interval}=[0.13046,0.29123].
\end{equation}

We denote the limit \(\mu=0\) by \(\Theta=\infty\), with the pair profile and
theory interpreted as \(\Theta\to\infty\). The interval above is the
Bonferroni adjusted 95\% family interval over the four tested modes. The
corresponding theory values for global relocation, no replacement, and their
difference were respectively -0.09351, 0.10894, and 0.20245. The two stages
used independent bootstraps and are supportive rather than part of
the primary analysis of the three replacement rules. It motivated separating
the removal of incident edges from density relocation in the present work.

\section{Long time dynamics under global relocation}

A separate nonlinear study followed trajectories with global relocation at
\(\eta=6\), \(H=2\), \(\Theta=5\), and \(\mathrm{Pe}=11\) to
dimensionless final time \(T=700\) for \(N=512,1024,2048\), with eight
independent seeds per size. To distinguish the observation windows from the
relocation radius \(R\), denote them by
\(W_1=(340,460]\), \(W_2=(460,580]\), and \(W_3=(580,700]\). This study
addresses the late nonlinear state; it does not enter the inference on early
mode growth above.

For completeness, let
\(c_{\boldsymbol m}(t)=N^{-1}\sum_{j=1}^N
e^{-\mathrm i\boldsymbol Q_{\boldsymbol m}\cdot X_j(t)}\) for
\(\boldsymbol m\in\mathbb Z^2\setminus\{\boldsymbol0\}\); thus
\(c_{\boldsymbol m}(t)\in\mathbb C\). For \(Q>0\), define the shell
\(\mathcal S_Q=\{\boldsymbol m:Q_{\boldsymbol m}=Q\}\), where
\(|\mathcal S_Q|\) is its number of modes. We restrict to allowed radii with
\(\mathcal S_Q\ne\varnothing\). The structure factor averaged over that shell is

\begin{equation*}
S(Q,t)=\frac{N}{|\mathcal S_Q|}
\sum_{\boldsymbol m\in\mathcal S_Q}|c_{\boldsymbol m}(t)|^2,
\qquad 0\leq \frac{S(Q,t)}{N}\leq1.
\end{equation*}

The smallest nonzero wavenumber is \(Q_{\min}=2\pi/L\). Values reported for
\(S(Q_{\min})/N\) are averages over the stated time window and simulation
seeds. For the tested band \(0<u\leq3\), the positive spectral excess weights
each shell by its number of modes and by \(\max[S(Q)-1,0]\). Its total is
\(E>0\); after normalization to unit mass, its first moment is the spectral
centroid \(\bar u>0\), its standard deviation is the spectral width, and
\(R_{0.75}\in[0,1]\) is the fraction below \(u=0.75\). We also record the
structure factor at one peak shell fixed for each system size over the full
observation interval. A core agent has contact degree at or above the smallest
integer whose upper tail probability under
\(\operatorname{Binomial}(N-1,\eta/(N-1))\) is at most 0.01. Here contact
degree is computed in the geometric graph with an edge whenever
\(r_{ij}\leq\sigma\), independently of persistent edge state; core clusters
are connected components of the subgraph induced by core agents.

The nine trajectory metrics used for autocorrelation checks were
\(\bar u\), spectral width, \(E\), \(R_{0.75}\),
\(S(Q_{\min})/N\), the fixed peak shell structure factor, the number of core
clusters divided by \(N\), the largest core cluster fraction, and the second
largest geometric contact component fraction. Temporal comparisons used the
same families except that the first Wasserstein distance between two normalized
excess spectra replaced the fixed peak shell amplitude.

All 216 combinations of trajectory and metric yielded valid integrated
autocorrelation estimates and covered at least 25 correlation times; the
minimum coverage was 36.47. The six consistency checks across system sizes
required relative spans below 5\% for \(\bar u\), 10\% for spectral width, and
10\% for the core cluster count per agent; nonincreasing largest core fraction
and \(S(Q_{\min})/N\); and a centroid on the largest system at least three
fundamental modes above zero. All six were satisfied. The late spectral
centroid, spectral width, and core cluster count
per agent had relative spans of 0.803\%, 2.103\%, and 0.232\%, respectively. The
largest core fraction
decreased from 0.0889 to 0.0495 to 0.0272, and
\(S(Q_{\min})/N\) decreased from 0.003750 to 0.001381 to 0.000596 as \(N\)
increased.

The stronger criterion for temporal stationarity was nevertheless not met.
The eight metric families other than the lowest Fourier shell satisfied all 72
temporal comparisons, whereas the quantity \(S(Q_{\min})/N\) at the lowest
shell satisfied four of nine. The only point estimate outside the
multiplicative equivalence band occurred for
\(N=512\), \(W_3/W_2=0.863520\), with 90\% interval
[0.815636,0.920231], compared with the lower boundary 0.869565. It followed an
increase in \(W_2/W_1\), while the comparison over the full observation span,
\(W_3/W_1=0.963757\), satisfied the criterion.

Values are log ratios; the common bounds
are \([-0.139762,0.139762]\).

\begin{table}[htbp]
\centering
\caption{Comparisons at the lowest Fourier shell that did not satisfy the prespecified equivalence criterion.}
\small
\setlength{\tabcolsep}{4pt}
\renewcommand{\arraystretch}{1.12}
\resizebox{\textwidth}{!}{%
\begin{tabular}{lcccc}
\toprule
\textbf{N} & \textbf{Comparison} & \textbf{Estimate} & \textbf{90\% interval} & \textbf{Reason not met} \\
\midrule
512 & W2/W1 & 0.109822 & [0.031439, 0.197641] & Upper interval endpoint outside \\
512 & W3/W2 & -0.146738 & [-0.203786, -0.083131] & Estimate and lower endpoint outside \\
1024 & W2/W1 & -0.082374 & [-0.187394, 0.019789] & Lower interval endpoint outside \\
1024 & W3/W2 & 0.063613 & [-0.064010, 0.182099] & Upper interval endpoint outside \\
2048 & W3/W2 & -0.088253 & [-0.176201, 0.003433] & Lower interval endpoint outside \\
\bottomrule
\end{tabular}%
}
\end{table}

The remaining four comparisons at the lowest shell satisfied the criterion,
including all three \(W_3/W_1\) comparisons over the full observation span.
Thus the unmet comparisons do not form a common monotone trend across size or
over the full observation span.

These data support a broad clustered state with a characteristic nonzero
wavenumber that is consistent across the tested sizes and times. They do not
establish a statistically stationary pattern. The lowest Fourier shell
displays slow, nonmonotone modulation rather than a common monotone drift
toward zero wavenumber.

\begin{figure}[htbp]
  \centering
  \includegraphics[width=0.96\textwidth]{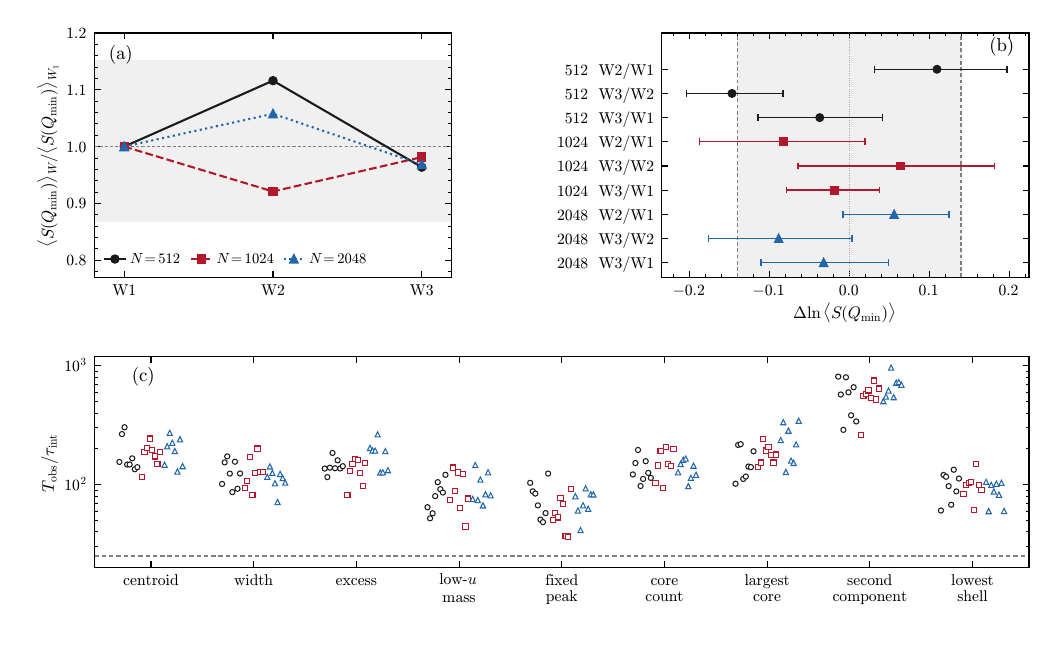}
  \caption{\textbf{Long time analysis of the lowest Fourier shell and system size effects.}
  Black circles, red squares, and blue triangles denote $N=512$, $1024$,
  and $2048$. (a) Mean lowest-shell structure factor in each late window,
  divided by its mean in $W_1$. The gray band is $[1/1.15,1.15]$;
  $W_1$, $W_2$, and $W_3$ are time windows, not displacement radii.
  (b) All nine log ratios of these window means, with their 90\% intervals.
  Row labels identify system size and window pair. The gray region and
  dashed boundaries mark $\pm\ln(1.15)$; an interval must lie entirely
  inside it to meet the temporal equivalence criterion.
  (c) Observation duration $T_{\rm obs}=360$ divided by the estimated
  integrated autocorrelation time $\tau_{\rm int}>0$, on a logarithmic
  vertical axis. Each metric has 24 points, one per independent trajectory
  at the three sizes, offset horizontally for visibility. The nine metrics
  are defined in the text; the dashed line is the required coverage of 25.
  The counts for the combined criteria are reported in
  Table~\ref{tab:longtime_criteria}.}
  \label{fig:supp-long}
\end{figure}

\begin{table}[htbp]
\centering
\caption{Components of the earlier long time assessment. These archived
checks concern the nonlinear late state and are separate from the primary
early-response criterion. Meeting the record-length and system-size checks
does not compensate for unresolved temporal equivalence.}
\label{tab:longtime_criteria}
\begin{tabular}{lrr}
\toprule
Criterion & Satisfied & Total \\
\midrule
Lowest-shell temporal equivalence & 4 & 9 \\
All temporal comparisons & 76 & 81 \\
Record-length sufficiency & 216 & 216 \\
Consistency across system sizes & 6 & 6 \\
Combined stationarity criterion & 0 & 1 \\
\bottomrule
\end{tabular}
\end{table}

\section{Integrity and reproducibility}

The archived prospective protocol, power calculation, prespecified configuration table
containing 76,800 rows, simulator, theory implementation, analysis code, and
tests were fixed
before the main simulations began.

For this primary experiment, the archived runtime record lists Python
3.13.14, NumPy 2.4.6, SciPy 1.18.0, and pandas 3.0.5 on arm64 macOS.
Independent verification covered all 76,800 trajectories, configuration
identifiers, time grids with 61 points, and common random number blocks
before the primary analysis. A separate
check reconstructed all raw and control variate point estimates from the
trajectories and regenerated the 10,000 bootstrap count vectors using analysis
seed \nolinkurl{2026081317}.

The independent supplemental studies contain 800 radius-scan blocks and
400 numerical-check blocks, totaling 25,600 trajectories. All completed
bundles, time grids, and disjoint seed sets were verified before analysis.
Each study uses 10,000 block-bootstrap draws with analysis seed
\nolinkurl{2026100209}. Their runtime records list Python 3.13.14, NumPy
2.3.5, and SciPy 1.18.0; these records are distinct from the archived
primary runtime above. Independent checks reconstruct the new rate
estimates, joint intervals, and local-structure summaries. The retrospective
window diagnostics use only the original trajectories and retain their
original block identities.

The additional mode and doubled-amplitude studies contain 3,200 and 1,600
new independent blocks, respectively, totaling 128,000 trajectories.
Their 4,800 seeds are unique and disjoint from all 9,116 historical and
reserved pilot seeds checked at protocol freeze. Each design was fixed
before production. Source versions, all block configurations, trajectory
counts, and recorded time grids were validated before outcome analysis.
The analysis seeds are \nolinkurl{2026100701} and
\nolinkurl{2026100702}, each with 10,000 whole-block resamples.
The paired radial reconstruction resamples its designated 200 blocks with
seed \nolinkurl{2026100713}. These additions neither pool the original
replicates nor alter its joint criterion. Complete tables, source manifests,
and independent numerical checks are archived separately for each design.

The retrospective early-window and rule comparisons were frozen on
12 October 2026 before their new computation. They use 10,000 complete-block
resamples with seeds \nolinkurl{2026101201}, \nolinkurl{2026101202}, and
\nolinkurl{2026101203} for the original, additional mode, and radius cohorts.
Independent reconstruction checked all 89 point fits and 356 selected
bootstrap fits with a separate scalar optimizer, rebuilt all seven
simultaneous families, and verified every reported count. Source and data
manifests, prior-data exposure, frozen expectations, and full numerical
precision are retained with this reanalysis.

The complete tables and analysis scripts are retained with the project and are
available from the corresponding author upon reasonable request.